\documentclass[vecphys]{svmult}

\usepackage{color}           
\usepackage{makeidx}         
\usepackage{graphicx}        
\usepackage{multicol}        
\usepackage{cite}            
\usepackage[bottom]{footmisc}

\makeindex             

\begin{document}

\title{Variational wave functions for frustrated magnetic models}
\author{Federico Becca,\inst{1} Luca Capriotti,\inst{2} Alberto Parola,\inst{3}
and Sandro Sorella\inst{1}}
\institute{CNR-INFM-Democritos National Simulation Centre and International 
School for Advanced Studies (SISSA), Via Beirut 2-4, I-34014 Trieste, Italy
\and Global Modelling and Analytics Group, Investment Banking Division, Credit
Suisse Group, Eleven Madison Avenue, NY-10010-3629 New York, United States
\and Dipartimento di Fisica e Matematica, Universit\`a dell'Insubria,
Via Valleggio 11, I-22100 Como, Italy}

\maketitle

Variational wave functions containing electronic pairing and suppressed 
charge fluctuations (i.e., projected BCS states) have been proposed as the 
paradigm for disordered magnetic systems (including spin liquids).
Here we discuss the general properties of these states in one and two
dimensions, and show that different quantum phases may be described with      
high accuracy by the same class of variational wave functions, including 
dimerized and magnetically ordered states. In particular, phases with magnetic
order may be obtained from a straightforward generalization containing both 
antiferromagnetic and superconducting order parameters, as well as suitable 
spin Jastrow correlations. In summary, projected wave functions represent an 
extremely flexible tool for understanding the physics of low-dimensional 
magnetic systems.

\section{Introduction}\label{sec:intro}

The variational approach is a widely used tool to investigate the low-energy 
properties of quantum systems with several active degrees of freedom, 
including electrons and ions. The basic idea is to construct fully quantum 
many-body states by a physically motivated \textit{ansatz}. The resulting 
wave function should be simple enough to allow efficient calculations even 
for large sizes. Most of the variational calculations are traditionally based 
upon mean-field approximations, where the many-body wave function is 
constructed by using independent single-particle states. In this respect, 
even the BCS theory of superconductivity belongs in this category~\cite{bcs}.
Although these mean-field approaches have been instrumental in understanding 
and describing weakly correlated systems, they have proved inadequate 
whenever the electron-electron interaction dominates the kinetic energy. The 
generalization of variational states in this regime is not straightforward, 
and represents an open problem in the modern theory of Condensed Matter. 
Probably the most celebrated case is the wave function proposed by Laughlin 
to describe the fractional quantum Hall effect as an incompressible quantum 
fluid with fractional excitations~\cite{laughlin}. One important example in 
which electron correlations prevent the use of simple, mean-field approaches 
is provided by the so-called resonating valence-bond (RVB) state. This 
intriguing phase, which was conjectured many years ago by Fazekas and 
Anderson~\cite{fazekas}, has no magnetic order, no broken lattice 
symmetries, and remains disordered even at zero temperature. It is now 
commonly accepted that these spin-liquid states may be stabilized in quantum 
antiferromagnets with competing (frustrating) interactions~\cite{misguich}. 

Here we present one possible approach to the definition of accurate variational 
wave functions which take into account quite readily both strong electron 
correlations and the frustrated nature of the lattice. The price to pay 
when considering these effects is that calculations cannot be performed 
analytically, and more sophisticated numerical methods, such as the quantum 
Monte Carlo technique, are required.

Let us begin by considering a simple, frustrated spin model, in which the 
combined effects of a small spin value, reduced dimensionality, and the 
presence of competing interactions could lead to non-magnetic phases. 
We consider what is known as the $J_1{-}J_2$ frustrated Heisenberg model
on a chain or a square lattice,
\begin{equation} \label{model}
{\cal H}=J_1\sum_{n.n.} {\bf S}_R \cdot {\bf S}_{R^\prime}
+ J_2\sum_{n.n.n.} {\bf {S}}_R \cdot {\bf {S}}_{R^\prime},
\end{equation}
where $J_1$ and $J_2$ are the (positive) nearest-neighbor ($n.n.$)
and next-nearest-neighbor ($n.n.n.$) couplings, and 
${\bf S}_R = (S^x_R,S^y_R,S^z_R)$ are $S=1/2$ operators; periodic boundary 
conditions are assumed. Besides the purely theoretical interest, this model 
is also known to describe the relevant antiferromagnetic interactions in a 
variety of quasi-one-dimensional~\cite{castilla} and quasi-two-dimensional 
systems~\cite{carretta,carretta2}.

In one dimension, the phase diagram of the $J_1{-}J_2$ model has been well 
established by analytical studies and by Density Matrix Renormalization Group 
(DMRG) calculations~\cite{white}. For small values of the ratio $J_2/J_1$, 
the system is in a Luttinger spin-fluid phase with a gapless spectrum, no 
broken symmetry, and power-law spin correlations. By increasing the value 
of the second-neighbor coupling, a gapped phase is 
stabilized~\cite{castilla,white}. The value of the critical point 
has been determined with high accuracy as $(J_2/J_1)_c = 0.241167 \pm 
0.000005$~\cite{eggert}. The gapped ground state is two-fold degenerate 
and spontaneously dimerized, and at $J_2/J_1 = 0.5$ is expressed by the 
exact Majumdar-Ghosh wave function~\cite{majumdar1,majumdar2}. 
Interestingly, for $J_2/J_1 > 0.5$, incommensurate but 
short-range spin correlations have been found, whereas the dimer-dimer 
correlations are always commensurate~\cite{white}.

By contrast, the phase diagram of the two dimensional $J_1{-}J_2$ model is 
the subject of much debate. For $J_2/J_1 \ll 0.5$, an antiferromagnetic 
N\'eel order with magnetic wave vector $Q=(\pi,\pi)$ is expected. In the 
opposite limit, $J_2/J_1 \gg 0.5$, the ground state is a collinear 
antiferromagnetic phase where the spins are aligned ferromagnetically in 
one direction and antiferromagnetically in the other [$Q=(\pi,0)$ or 
$Q=(0,\pi)$]. The nature of the ground state in the regime of strong 
frustration, i.e., for $J_2/J_1\sim 0.5$, remains an open problem, 
and there is no general consensus on its characterization. Since the 
work of Chandra and Doucot~\cite{doucot}, it has been suggested that a 
non-magnetic phase should be present around $J_2/J_1 = 0.5$. Unfortunately, 
exact diagonalization calculations are limited to small clusters which 
cannot provide definitive answers to this very delicate 
problem~\cite{dagotto,singh,schulz}. By using series-expansion 
methods~\cite{gelfand,singh2,kotov,sushkov} and field-theoretical 
approaches~\cite{read}, it has been argued that a valence-bond solid, 
with columnar dimer order and spontaneous symmetry-breaking, could be 
stabilized. More recently, it has been 
shown that a clear enhancement of plaquette-plaquette correlations is 
found by introducing a further, third-nearest-neighbor superexchange 
term $J_3$, thus suggesting a possible plaquette valence-bond 
crystal~\cite{mambrini}. 

The primary obstacle to the characterization of the phase diagram in two 
dimensions is that the lack of exact results is accompanied, in the frustrated 
case, by difficulties in applying standard stochastic numerical techniques. 
Quantum Monte Carlo methods can be applied straightforwardly only to spin-1/2 
Hamiltonians of the form~(\ref{model}), with strong restrictions on the 
couplings (e.g., $J_1 \ge  0$ and $J_2 \le 0$ or $J_1 \le 0$ and $J_2 
\le 0$) in order to avoid a numerical instability known as the \textit{sign 
problem}. This is because, in general, quantum Monte Carlo methods do not 
suffer from numerical instabilities only when it is possible to work with a 
basis in the Hilbert space where the off-diagonal matrix elements of the 
Hamiltonian are all non-positive. As an example, in a quantum antiferromagnet 
with $J_1\ge 0$ and $J_2=0$ on a bipartite lattice, after the unitary 
transformation
\begin{equation}
{\cal U}^\dagger = 
\exp{\Big[-i \pi \sum_{R \in {\rm B}} S_R^z \Big]}
\label{eq.rotpi}
\end{equation}
(${\rm B}$ being one of the two sublattices), the transformed Hamiltonian
has non-positive off-diagonal matrix elements in the basis $|x\rangle$
whose states are specified by the value of $S_R^z$ on each site, and 
$\sum_R S_R^z=S$~\cite{lieb}. This implies that the ground state of 
$\cal{U} \cal{H} \cal{U}^\dagger$, 
$|\tilde{\Psi}_0\rangle = \sum_x \tilde{\Psi}_0(x)|x\rangle$, 
has all-positive amplitudes, $\tilde{\Psi}_0(x) > 0$, 
meaning that there exists a purely bosonic representation of the ground 
state. This property leads to the well-known Marshall-Peierls sign 
rule~\cite{lieb,marshall} for the phases of the ground state of ${\cal H}$, 
${\rm sign} \{\Psi_0(x)\}=(-1)^{N_\uparrow(x)}$, where $N_\uparrow(x)$ is
the number of up spins on one of the two sublattices. The Marshall-Peierls 
sign rule holds for the unfrustrated Heisenberg model and even for the 
$J_1{-}J_2$ chain at the Majumdar-Ghosh point. However, in 
the regime of strong frustration, the Marshall-Peierls sign rule is violated 
dramatically~\cite{richter}, and, because no analogous sign rule appears 
to exist, the ground-state wave function has non-trivial 
phases. This property turns out to be a crucial ingredient of frustration.

In this respect, a very useful way to investigate the highly frustrated 
regime is to consider variational wave functions, whose accuracy can be
assessed by employing stable (but approximate) Monte Carlo techniques 
such as the fixed-node approach~\cite{ceperley}.
Variational wave functions can be very flexible, allowing the description of 
magnetically ordered, dimerized, and spin-liquid states. In particular, it is 
possible to construct variational states with non-trivial signs for the 
investigation of the strongly frustrated regime.

In the following, we will describe in detail the case in which the variational 
wave function is constructed by projecting \textit{fermionic} mean-field 
states~\cite{anderson2}. Variational calculations can be 
treated easily by using standard Monte Carlo techniques. This is in contrast 
to variational states based on a \textit{bosonic} representation, which
are very difficult to handle whenever the ground state has non-trivial 
phases~\cite{liang}. 
Indeed, variational Monte Carlo calculations based on bosonic wave functions 
suffer from the sign problem in the presence of frustration~\cite{sindzingre}, 
and stable numerical simulations can be performed only in special cases, for 
example in bipartite lattices when the valence bonds only connect opposite 
sublattices~\cite{liang}. 
Another advantage of the fermionic representation is that the mean-field 
Hamiltonian allows one to have a simple and straightforward representation 
also for the low-lying excited states (see the discussion in 
section~\ref{sec:oned}, and also Ref.~\cite{becca} for a frustrated model 
on a three-leg ladder). 

\section{Symmetries of the wave function: general properties}\label{sec:symm}

We define the class of projected-BCS (pBCS) wave functions on an $N$-site 
lattice, starting from the ground state of a suitable translationally 
invariant BCS Hamiltonian  
\begin{eqnarray}
{\cal H}_{BCS}&=&
\sum_{R,R^\prime \sigma} (t_{R-R^\prime}-\mu\,\delta_{R-R^\prime})\, 
c^\dagger_{R,\sigma}c_{R^\prime,\sigma}
-\sum_{R,R^\prime} \Delta_{R-R^\prime} \,
c^\dagger_{R,\uparrow}c^\dagger_{R^\prime,\downarrow} + H.c. \nonumber\\
&=&
\sum_{k \sigma} (\epsilon_{k}-\mu)\, c^\dagger_{k,\sigma}c_{k,\sigma}
-\sum_{k} \Delta_{k} c^\dagger_{k,\uparrow}c^\dagger_{-k,\downarrow} + H.c.,
\label{hbcs}
\end{eqnarray}
where $c^\dagger_{R,\sigma}$ ($c_{R,\sigma}$) creates (destroys) an electron
at site $R$ with spin $\sigma$, the bare electron band $\epsilon_k$ is a real 
and even function of $k$, and $\Delta_k$ is also taken to be even to describe 
singlet electron pairing. In order to obtain a class of non-magnetic, 
translationally invariant, and singlet wave functions for spin-1/2 models, 
the ground state $|BCS \rangle$ of Hamiltonian~(\ref{hbcs}) is projected 
onto the physical Hilbert space of singly-occupied sites by the Gutzwiller 
operator $P_G = \prod_R (n_{R,\uparrow}-n_{R,\downarrow})^2$, $n_{R,\sigma}$ 
being the local density. Thus 
\begin{equation}\label{bcs2}
|pBCS \rangle = P_G \, |BCS \rangle =
P_G \prod_k (u_k+v_k c^\dagger_{k,\uparrow}c^\dagger_{-k,\downarrow}) 
|0\rangle,
\end{equation}
where the product is over all the $N$ wave vectors in the Brillouin zone.
The diagonalization of Hamiltonian~(\ref{hbcs}) gives explicitly
$$
u_k = \sqrt{\frac{ E_k+\epsilon_k}{2 E_k} }  \qquad\qquad
v_k = \frac{\Delta_k}{|\Delta_k |} \sqrt{\frac{ E_k-
\epsilon_k}{2 E_k} }  \qquad\qquad
E_k = \sqrt{ \epsilon_k^2 +|\Delta_k|^2 },
$$
while the BCS pairing function $f_k$ is given by
\begin{equation}\label{fk}
f_k = \frac{v_k}{u_k} = \frac{\Delta_k}{\epsilon_k + E_k}.
\end{equation}

The first feature we wish to discuss is the \textit{redundancy} implied by 
the electronic representation of a spin state, by which is meant the extra 
symmetries which appear when we write a spin state as the Gutzwiller 
projection of a fermionic state. This property is reflected in turn in 
the presence of a \textit{local} gauge symmetry of the fermionic 
problem~\cite{affleck,rice,wen}. Indeed, the original spin 
Hamiltonian~(\ref{model}) is invariant under the local SU(2) gauge 
transformations
\begin{eqnarray}
&&\Sigma^z_\phi:
\left (\begin{array}{c} c^\dag_\uparrow \cr c_\downarrow \end{array}\right )
\to 
e^{i \phi {\bf \sigma}_z}
\left ( 
\begin{array}{c} c^\dag_\uparrow \cr c_\downarrow 
\end{array}
\right ) =
\left ( 
\begin{array}{cc} 
e^{i \phi} & 0 \cr
0           & e^{-i \phi} \end{array} \right )
\left ( 
\begin{array}{c} c^\dag_\uparrow \cr c_\downarrow 
\end{array}
\right ),
\label{u1} \\
&&\Sigma^x_\theta:
\left (\begin{array}{c} c^\dag_\uparrow \cr c_\downarrow \end{array}\right )
\to 
e^{i \theta {\bf \sigma}_x}
\left ( 
\begin{array}{c} c^\dag_\uparrow \cr c_\downarrow 
\end{array}
\right ) =
\left ( 
\begin{array}{cc} 
\cos\theta & i \sin\theta \cr
i\sin\theta &\cos\theta \end{array} \right )
\left ( 
\begin{array}{c} c^\dag_\uparrow \cr c_\downarrow 
\end{array}
\right ).
\label{sigma}
\end{eqnarray}
A third transformation can be expressed in terms of the previous ones,
\begin{equation}
\Sigma^y_\psi:
\left (\begin{array}{c} c^\dag_\uparrow \cr c_\downarrow \end{array}\right )
\to 
e^{i \psi {\bf \sigma}_y}
\left ( 
\begin{array}{c} c^\dag_\uparrow \cr c_\downarrow 
\end{array}
\right ) =
e^{-i \pi {\bf \sigma}_z/4} e^{i \psi {\bf \sigma}_x} e^{i \pi {\bf \sigma}_z/4}
\left ( 
\begin{array}{c} c^\dag_\uparrow \cr c_\downarrow 
\end{array}
\right ),
\label{third}
\end{equation}
where ${\bf \sigma}_x$, ${\bf \sigma}_y$, and ${\bf \sigma}_z$ are the Pauli 
matrices. As a consequence, all the different fermionic states connected by a 
local SU(2) transformation generated by~(\ref{u1}) and~(\ref{sigma}) 
with site-dependent parameters give rise to the same spin state after 
Gutzwiller projection,  
\begin{equation}
P_G\,\prod_R \Sigma^z_{\phi_R} \Sigma^x_{\theta_R} \,|BCS \rangle = 
e^{i\Phi} P_G\,|BCS \rangle,
\end{equation}
where $\Phi$ is an overall phase. Clearly, the local gauge transformations 
defined previously change the BCS Hamiltonian, breaking in general the 
translational invariance. In the following, we will restrict our 
considerations to the class of transformations which preserve the 
translational symmetry of the lattice in the BCS Hamiltonian, i.e., 
the subgroup of global symmetries corresponding to site-independent angles 
$(\phi,\theta)$. By applying the transformations~(\ref{u1}) and~(\ref{sigma}), 
the BCS Hamiltonian retains its form with modified couplings
\begin{eqnarray}
t_{R-R^\prime}&\to& t_{R-R^\prime} \\
\Delta_{R-R^\prime}&\to&\Delta_{R-R^\prime} e^{2i\phi}
\end{eqnarray}
for $\Sigma^z_\phi$, while the transformation $\Sigma^x_\theta$ gives
\begin{eqnarray}
t_{R-R^\prime}&\to& \cos 2\theta \, t_{R-R^\prime} +i\sin \theta\cos\theta \, 
(\Delta_{R-R^\prime} -\Delta^*_{R-R^\prime}) \nonumber\\
&=&\cos 2\theta \, t_{R-R^\prime}-\sin 2\theta \, {\rm Im}\Delta_{R-R^\prime} \\
\Delta_{R-R^\prime}&\to& (\cos^2 \theta \, \Delta_{R-R^\prime} + \sin^2\theta \,
\Delta^*_{R-R^\prime})+ i\sin 2\theta \, t_{R-R^\prime} \nonumber\\
&=&{\rm Re} \Delta_{R-R^\prime} + i\left ( \cos 2\theta \, 
{\rm Im}\Delta_{R-R^\prime}+\sin 2\theta \, t_{R-R^\prime} \right ).
\end{eqnarray}
These relations are linear in $t_{R-R^\prime}$ and $\Delta_{R-R^\prime}$, 
and therefore hold equally for the Fourier components $\epsilon_k$ and 
$\Delta_k$. We note that, because $\Delta_r$ is an even function, the real 
(imaginary) part of its Fourier transform $\Delta_k$ is equal to the Fourier 
transform of the real (imaginary) part of $\Delta_r$. It is easy to see that 
these two transformations generate the full rotation group on the vector 
whose components are $(\epsilon_k, \,{\rm Re} \Delta_k,\,{\rm Im}\Delta_k)$.
As a consequence, the length $E_k$ of this vector is conserved by the full 
group. 

In summary, there is an infinite number of different translationally invariant 
BCS Hamiltonians that, after projection, give the same spin state. Choosing a 
specific representation does not affect the physics of the state, but changes 
the pairing function $f_k$ of Eq.~(\ref{fk}) before projection. 
Within this class of states, the only scalar under rotations is the BCS energy 
spectrum $E_k$. Clearly, the projection operator will modify the excitation 
spectrum associated with the BCS wave function. Nevertheless, its invariance 
with respect to SU(2) transformations suggests that $E_k$ may reflect the 
nature of the physical excitation spectrum.

Remarkably, in one dimension it is easy to prove that such a class of wave 
functions is able to represent faithfully both the physics of Luttinger 
liquids, appropriate for the nearest-neighbor Heisenberg model, and the 
gapped spin-Peierls state, which is stabilized for sufficiently strong 
frustration. In fact, it is known~\cite{haldane} that the simple choice 
of nearest-neighbor hopping ($\epsilon_k=-2t\,\cos k$, $\mu=0$) and vanishing 
gap function $\Delta_k$ reproduces the exact solution of the Haldane-Shastry 
model (with a gapless $E_k$), while choosing a next-nearest neighbor hopping 
($\epsilon_k=-2t\,\cos 2k$, $\mu=0$) and a sizable nearest-neighbor pairing 
($\Delta_k=4\sqrt{2}t\,\cos k$) recovers the Majumdar-Ghosh state (with a 
gapped $E_k$)~\cite{majumdar1,majumdar2}.

\section{Symmetries in the two-dimensional case}\label{sec:symm2d}

We now specialize to the two-dimensional square lattice and investigate 
whether it is possible to exploit further the redundancy in the fermionic 
representation of a spin state in order to define a pairing function which
breaks some spatial symmetry of the lattice but which, after projection, 
still gives a wave function with all of the correct quantum numbers. We 
will show that, if suitable conditions are satisfied, a fully symmetric 
projected BCS state is obtained from a BCS Hamiltonian with fewer 
symmetries than the original spin problem. For this purpose, it is 
convenient to introduce a set of unitary operators related to the 
symmetries of the model.
\begin{itemize}
\item
{Spatial symmetries: for example, ${\cal R}_x (x,y)=(x,-y)$ and
${\cal R}_{xy}(x,y)=(y,x)$. We define the transformation law of creation 
operators as  
${\cal R} c^\dagger_{X,\sigma} {\cal R}^{-1} = c^\dagger_{{\cal R}(X),\sigma}$,
and the action of the symmmetry operator on the vacuum is  
${\cal R}_x |0\rangle={\cal R}_{xy} |0\rangle=|0\rangle$.
Note that these operators map each sublattice onto itself.}

\item
{Particle-hole symmetry: 
$P_h c^\dagger_{X,\sigma} P_h^{-1} = i\,(-1)^X c_{X,-\sigma}$, 
where the action of the $P_h$ operator on the vacuum is 
$P_h|0\rangle=\prod_X c^\dagger_{X,\uparrow} c^\dagger_{X,\downarrow} 
|0\rangle$.}

\item
{Gauge transformation: $G\,c^\dagger_{X,\sigma}G^{-1}=i\,c^\dagger_{X,\sigma}$
with $G\,|0\rangle=|0\rangle$.}
\end{itemize}

Clearly, ${\cal R}_x$ and ${\cal R}_{xy}$ are symmetries of the physical 
problem (e.g., the Heisenberg model). $G$ is a symmetry because the 
physical Hamiltonian has a definite number of electrons, while $P_h$ leaves 
invariant every configuration where each site is singly occupied if the total 
magnetization vanishes ($N_\downarrow= N_\uparrow=n$). With the definition 
adopted, $P_h$ acts only to multiply every spin state by the phase factor 
$(-1)^{N_\downarrow}$. Thus all of the operators defined above commute both 
with the Heisenberg Hamiltonian and, because reflections do not interchange 
the two sublattices, with each other. The ground state of the Heisenberg 
model on a finite lattice, if it is unique, must be a simultaneous eigenstate 
of all the symmetry operators. We will establish the sufficient conditions 
which guarantee that the projected BCS state is indeed an eigenstate of all 
of these symmetries.

Let us consider a hopping term which only connects sites in opposite 
sublattices, whence $\epsilon_{k+Q}=-\epsilon_k$, and a gap function with 
contributions from different symmetries ($s$, $d_{x^2-y^2}$, and $d_{xy}$), 
$\Delta=\Delta^s+\Delta^{x^2-y^2}+ \Delta^{xy}$. Further, we consider a 
case in which $\Delta^s$ and $\Delta^{x^2-y^2}$ couple opposite sublattices, 
while $\Delta^{xy}$ is restricted to the same sublattice. In this case, the 
BCS Hamiltonian 
${\cal H}_{BCS} = {\cal H} (t, \Delta^s, \Delta^{x^2-y^2}, \Delta^{xy})$ 
transforms under the different unitary operators acording to 
\begin{eqnarray}
{\cal R}_x {\cal H}(t,\Delta^s,\Delta^{x^2-y^2},\Delta^{xy}) 
{\cal R}_x^{-1} &=& {\cal H}(t,\Delta^s,\Delta^{x^2-y^2},-\Delta^{xy}), 
\nonumber \\
{\cal R}_{xy} {\cal H}(t,\Delta^s,\Delta^{x^2-y^2},\Delta^{xy}) 
{\cal R}_{xy}^{-1} &=& {\cal H}(t,\Delta^s,-\Delta^{x^2-y^2},\Delta^{xy}), 
\nonumber \\
P_h {\cal H}(t,\Delta^s,\Delta^{x^2-y^2},\Delta^{xy}) P_h^{-1} &=&
{\cal H}(t,{\Delta^s}^*,{\Delta^{x^2-y^2}}^*,-{\Delta^{xy}}^*), \nonumber \\
G {\cal H}(t,\Delta^s,\Delta^{x^2-y^2},\Delta^{xy}) G^{-1} &=&
{\cal H}(t,-\Delta^s,-\Delta^{x^2-y^2},-\Delta^{xy}). \nonumber
\end{eqnarray}
From these transformations it is straightforward to define suitable 
composite symmetry operators which leave the BCS Hamiltonian invariant. 
For illustration, in the case where $\Delta$ is real, one may select 
${\cal R}_xP_h$ and ${\cal R}_{xy}$ if $\Delta^{x^2-y^2}=0$ or 
${\cal R}_xP_h$ and ${\cal R}_{xy}P_h G$ if $\Delta^s=0$. It is not 
possible to set both $\Delta^{x^2-y^2}$ and $\Delta^s$ simultaneously 
different from zero and still obtain a state with all the symmetries of 
the original problem. The eigenstates $|BCS\rangle$ of Eq.~(\ref{hbcs}) 
will in general be simultaneous eigenstates of these two composite 
symmetry operators with given quantum numbers, for example $\alpha_x$ 
and $\alpha_{xy}$. The effect of projection over these states is
\begin{eqnarray}\label{eigenx}
\alpha_x\, P_G |BCS\rangle &=& P_G {\cal R}_x P_h |BCS\rangle
= {\cal R}_x P_h P_G |BCS\rangle \nonumber \\
&=& (-1)^{n}{\cal R}_x P_G |BCS\rangle,
\end{eqnarray}
where we have used that both ${\cal R}_x$ and $P_h$ commute with the 
projector. Analogously, when a term $\Delta_{x^2-y^2}$ is present,
\begin{eqnarray}\label{eigenxy}
\alpha_{xy}\, P_G |BCS\rangle &=& P_G {\cal R}_{xy} P_h G|BCS\rangle =
{\cal R}_{xy} P_h G P_G |BCS\rangle \nonumber \\
&=& {\cal R}_{xy} P_G |BCS\rangle.
\end{eqnarray}
These equations show that the projected BCS state with both $s$ and $d_{xy}$, 
or $d_{x^2-y^2}$ and $d_{xy}$, contributions to the gap has definite symmetry 
under reflections, in addition to being translationally invariant. The 
corresponding eigenvalues, for $n=N/2$ even, coincide with the eigenvalues 
of the modified symmetry operators ${\cal R}_x P_h$ and ${\cal R}_{xy} P_h G$ 
on the pure BCS state. 

In the previous discussion of quantum numbers, it was assumed that
$u_k$ and $v_k$ are well defined for every wave vector $k$. However, 
this condition is in general violated: singular $k$-points are present 
whenever both the band structure $\epsilon_k$ and the gap function $\Delta_k$ 
vanish, as for example with $\mu=0$, nearest-neighbor hopping and 
$d_{x^2-y^2}$ pairing at $k=(\pm\frac{\pi}{2},\pm\frac{\pi}{2})$.
However, on finite lattices, this occurrence can be avoided by the choice 
of suitable boundary conditions. In fact we are free to impose either 
periodic or antiperiodic boundary conditions on the fermionic BCS 
Hamiltonian~(\ref{hbcs}), while maintaining all the symmetries of the original 
lattice. In our studies we have selected lattices and boundary conditions 
which do not result in singular $k$-points. We note that the quantum numbers 
of the projected state do depend in general on the choice of boundary 
conditions in the fermionic BCS Hamiltonian. 

\subsection{The Marshall-Peierls sign rule}\label{subsec:marshall}

Another interesting property of the class of pBCS wave functions is related 
to the possibility of satisfying the Marshall-Peierls sign rule by means of 
a suitable choice of the gap function. In particular, we will restrict our 
considerations to the class of projected wave functions specified in 
Eq.~(\ref{bcs2}) when both $t_{R-R^\prime}$ and $\Delta_{R-R^\prime}$ 
are real and couple sites in opposing sublattices. We begin with the BCS 
Hamiltonian~(\ref{hbcs}) and perform a particle-hole transformation on 
the down spins alone, $d^\dagger_{R,\uparrow}=c^\dagger_{R,\uparrow}$ and
$d^\dagger_{R,\downarrow}=e^{iQ\cdot R} c_{R,\downarrow}$, with $Q=(\pi,\pi)$,
followed by the canonical transformation (spin rotation)
$a_+(k)=(d_{k,\uparrow}+id_{k,\downarrow})/\sqrt{2}$ and
$a_-(k)=-i(d_{k,\uparrow}-id_{k,\downarrow})/\sqrt{2}$. The BCS 
Hamiltonian then acquires the form
\begin{equation}
{\cal H}_{BCS}=\sum_k \left [h_+(k) + h_-(k) \right ],
\end{equation}
where $h_\pm (k) = \epsilon_k a_\pm^\dagger(k) a_\pm(k) \pm i\Delta_k
a_\pm^\dagger(k) a_\pm(k+Q)$, and we have used the symmetry $\Delta_k = 
-\Delta_{k+Q}$. Due to the anticommutation rules of the operators $a_\pm(k)$, 
the ground state of ${\cal H}_{BCS}$ can be written as a tensor product of 
free states of $\pm$ fermions. Moreover, if $\sum_R \Psi(R_1\cdots R_n) 
a_+^\dagger(R_1)\cdots a_+^\dagger(R_n) |0\rangle$ is the ground state 
of $\sum_k h_+(k)$, then $\sum_R \Psi^*(R_1\cdots R_n) a_-^\dagger(R_1) 
\cdots a_-^\dagger(R_n) |0\rangle$ is the ground state of $\sum_k h_-(k)$. 
Here we have chosen an arbitrary ordering of the lattice sites.
The ground state of ${\cal H}_{BCS}$ is therefore 
\begin{equation}
\sum_{R,R^\prime} \Psi(R_1\cdots R_n)\Psi^*(R^\prime_1 \cdots R^\prime_n)
a^\dagger_+(R_1)\cdots a^\dagger_+(R_n) a^\dagger_-(R^\prime_1)
\cdots a^\dagger_-(R^\prime_n) |0 \rangle.
\end{equation}
If this state is expressed in terms of the original electron operators we 
obtain, up to a factor of proportionality,
\begin{eqnarray}
&&\sum_{R,R^\prime} \Psi(R_1\cdots R_n)\Psi^*(R^\prime_1 \cdots R^\prime_n) 
\left [ c^\dagger_{R_1,\uparrow}- i e^{iQ\cdot R_1} c_{R_1,\downarrow} \right] 
\cdots \nonumber \\
&&\left [ c^\dagger_{R^\prime_1,\uparrow}+ i e^{iQ\cdot R^\prime_1} 
c_{R^\prime_1,\downarrow} \right] 
\cdots c^\dagger_{1,\downarrow} \cdots c^\dagger_{N,\downarrow} |0 \rangle.
\end{eqnarray}
In projecting over the state of fixed particle number equal to the number 
of sites, we must take the same number of creation and annihilation operators 
in the $N$ factors of the product. The suppression of doubly occupied sites 
mandated by the Gutzwiller projector is effected by creating an up spin on 
sites where a down spin has already been annihilated. The only terms which 
survive are then those with $\{R\}=\{R^\prime\}$, namely 
\begin{equation}
\sum_{R} |\Psi(R_1\cdots R_n)|^2 e^{i \sum_j Q \cdot R_j}
c^\dagger_{R_1,\uparrow} c_{R_1,\downarrow} \cdots
c^\dagger_{R_n,\uparrow} c_{R_n,\downarrow}
\cdots c^\dagger_{1,\downarrow} \cdots c^\dagger_{N,\downarrow}
|0 \rangle.
\end{equation}
Finally, by moving the down-spin creation operators to the left, one may
order the operators according to the specified ordering of the sites in the
lattice, independently of the spin, without introducing any further phase
factors. On this basis, the resulting wave function has exactly the 
Marshall-Peierls sign.

\subsection{Spin correlations}\label{subsec:spincorr} 

Finally, we would like to calculate the form of the long-range decay 
of the spin correlations in a BCS state. Here, we will show only that the 
pure BCS state before projection is characterized by correlations which 
maintain the symmetries of the lattice even when the BCS Hamiltonian breaks 
the reflection symmetries due to the presence of both $\Delta^{x^2-y^2}$ 
and $\Delta^{xy}$ couplings. Because the BCS state~(\ref{bcs2}) is a 
translationally invariant singlet, it is sufficient to calculate the 
longitudinal correlations $\langle S^z_r S^z_0 \rangle$. A straightforward 
application of Wick's theorem leads (for $r\ne 0$) to
\begin{eqnarray}
\label{corre1}
\langle S^z_r S^z_0 \rangle &\propto& -\left [ g^2(r)+h^2(r)\right ], \\
g(r)&=&\int d^2k\, {\epsilon_k \over E_k} \,e^{ik\cdot r}, \\
h(r)&=&\int d^2k \,{\Delta_k \over E_k}\, e^{ik\cdot r}.
\end{eqnarray}
Note that when the gap function $\Delta_k$ has both $d_{x^2-y^2}$ and 
$d_{xy}$ contributions, the correlation function apparently breaks rotational
invariance. Equation~(\ref{corre1}) can be written equivalently in Fourier 
space as
\begin{equation}
S(q)= \langle S^z_q S^z_{-q} \rangle \,\propto \int d^2k\, 
{\epsilon_k\epsilon_{k+q} + \Delta_k\Delta_{k+q} \over 
\sqrt{\Big [ \epsilon_k^2 +\Delta_k^2\Big ]
\,\,\left [\epsilon_{k+q}^2 +\Delta_{k+q}^2 \right ] }}.
\end{equation}
Now the effect of an $x$-reflection ${\cal R}_x$ on the wave vector $q$ 
can be deduced by setting $\Delta_k = \Delta^{x^2-y^2}_k + \Delta^{xy}_k$ 
and changing the dummy integration variable $k \to {\cal R}_x \,(k+Q)$, 
whence $\Delta_{k+Q}=-\Delta^{x^2-y^2}_k + \Delta^{xy}_k$ and
$\Delta_{R_xk}=\Delta^{x^2-y^2}_k - \Delta^{xy}_k$. The net result of 
these transformations is simply $S({\cal R}_x q)=S(q)$, demonstrating that 
the spin correlations of a BCS state are isotropic, even if the gap function 
breaks rotational invariance before Gutzwiller projection.

The explicit evaluation of the long-range decay of $g(r)$ for a $d_{x^2-y^2}$ 
gap shows that spin correlations in a BCS state (i.e., before projection) 
display a power-law decay due to the presence of gapless modes: 
$\langle S^z_r S^z_0 \rangle \sim 1/r^4$ for sites on opposite sublattices,
while $\langle S^z_r S^z_0 \rangle$ vanishes for sites on the same sublattice.
A similar result is also expected in the presence of a finite $\Delta^{xy}_k$,
because gapless modes are present also in this case.

\section{Connection with the bosonic representation}\label{sec:boson}

We turn now to a detailed discussion of the relation between the 
fermionic~\cite{anderson2} and bosonic~\cite{liang} representations of 
the RVB wave function. Recently, bosonic RVB wave functions have been 
reconsidered by Beach and Sandvik~\cite{lou,sandvik,sandvik2,beach}. 
In particular, it has been possible to improve the earlier results of 
Ref.~\cite{liang}, either by assuming some asymptotic form of the bond 
distribution~\cite{beach} or by unconstrained numerical methods~\cite{lou}. 
This wave function has been demonstrated to be extremely accurate for the
unfrustrated model with $J_2=0$~\cite{liang,lou}.

In the fermionic representation, we have
\begin{equation}
|pBCS \rangle = P_G \exp \left[ \sum_{R<R^\prime} f_{R,R^\prime} 
(c^\dagger_{R,\uparrow} c^\dagger_{R^\prime,\downarrow} +
c^\dagger_{R^\prime,\uparrow} c^\dagger_{R,\downarrow}) \right] |0\rangle,
\end{equation}
where $P_G$ projects onto the physical subspace with one electron per site
and $f_{R,R^\prime}$ is the pairing function, given by the Fourier transform 
of Eq.~(\ref{fk}). The constraint $R<R^\prime$ implies the definition of an 
(arbitrary) ordering of the lattice sites: here and in the following, we 
will refer to the lexicographical order. For simplicity, let 
us denote the singlet operator as
$\Theta_{R,R^\prime} =(c^\dagger_{R,\uparrow} c^\dagger_{R^\prime,\downarrow}
 + c^\dagger_{R^\prime,\uparrow} c^\dagger_{R,\downarrow})$.
Once the Gutzwiller projector is taken into account, we have that
\begin{equation}
|pBCS \rangle =
\sum_{ R_1 < \dots < R_n}\,\sum_{P(R^\prime)}
f_{R_1,R^\prime_1} \dots f_{R_n,R^\prime_n} 
\Theta_{R_1,R^\prime_1} \dots \Theta_{R_n,R^\prime_n}
|0\rangle,
\end{equation}
where $n=N/2$ and $P(R^\prime)$ represents the permutations of the $n$ sites 
$R^\prime_k$ not belonging to the set $\{R\}$, satisfying $R_k < R^\prime_k$ 
for every $k$. The sum defines all the $(N-1)!!$ partitions of the $N$ sites
into pairs.

On the other hand, the bosonic RVB wave function may be expressed in terms 
of the spin-lowering operator, $S^{-}_R$, as
\begin{equation}\label{bosonicrvb}
|RVB \rangle = 
\sum_{ R_1 < \dots < R_n }\,\sum_{P(R^\prime)}
f^{bos}_{R_1,R^\prime_1} \dots f^{bos}_{R_n,R^\prime_n} 
(S^{-}_{R^\prime_1} - S^{-}_{R_1}) \dots 
(S^{-}_{R^\prime_n} - S^{-}_{R_n})
|F\rangle,
\end{equation}
where the sum has the same restrictions as before and $|F\rangle$ is the 
(fully polarized) ferromagnetic state. In the bosonic representation, 
a valence-bond singlet is antisymmetric on interchanging the two sites and, 
therefore, a direction must be specified. The condition $R_k<R^\prime_k$ 
fixes the phase (i.e., the sign) of the RVB wave function.
In order to make contact between the two representations, we express 
$|F\rangle$ and $S^{-}_R$ in terms of fermionic operators, namely 
$|F\rangle = c^\dagger_{1,\uparrow}\dots c^\dagger_{N,\uparrow} |0\rangle$
and $S^{-}_R=c^\dagger_{R,\downarrow} c_{R,\uparrow}$. Then
\begin{equation}
|RVB \rangle =
\sum_{R_1 < \dots < R_n} \,\sum_{P(R^\prime)}\,\epsilon_{\{R,R^\prime\}}
f^{bos}_{R_1,R^\prime_1} \dots f^{bos}_{R_n,R^\prime_n}
\Theta_{R_1,R^\prime_1} \dots \Theta_{R_n,R^\prime_n} |0\rangle,
\end{equation}
where $\epsilon_{\{R,R^\prime\}}=\pm 1$ is a configuration-dependent 
sign arising from the reordering of the fermionic operators 
$(1,\dots N) \to (R_1,R^\prime_1 \dots R_n,R^\prime_n)$.
The two representations are therefore equivalent only if 
\begin{equation}\label{bos-fer}
\epsilon_{\{R,R^\prime\}} \,
f_{R_1,R^\prime_1} \dots f_{R_n,R^\prime_n} =
f^{bos}_{R_1,R^\prime_1} \dots f^{bos}_{R_n,R^\prime_n}
\end{equation}
for all the valence-bond configurations. In general, for a given 
$f^{bos}_{R,R^\prime}$, this condition cannot be satisfied by any choice of 
$f_{R,R^\prime}$. Remarkably, this is however possible for the short-range 
RVB state~\cite{sutherland,chakraborty}, where only nearest-neighbor sites 
are coupled by $f^{bos}_{R_k,R^\prime_k}=1$. Indeed, by using the Kasteleyn 
theorems~\cite{kasteleyn}, it is possible to prove that Eq.~(\ref{bos-fer}) 
can be fulfilled on all planar graphs (for example in short-range RVB states 
on lattices with open boundary conditions). In fact, the left-hand side of 
Eq.~(\ref{bos-fer}) is a generic term in the Pfaffian of the matrix
\begin{equation}
M(R,R^\prime)=\cases{ \;\; f_{R,R^\prime} & for $R<R^\prime$, \cr
                          -f_{R^\prime,R} & for $R>R^\prime$.}
\end{equation}
As a consequence, following the arguments of Kasteleyn, it is always possible 
to orient all the bonds in such a way that \textit{in all cycles of the 
transition graph the number of bonds oriented in either directions is 
odd}~\cite{kasteleyn}. Notice that the latter way to orient the bonds will 
in general be different from the one used in Eq.~(\ref{bosonicrvb}). Thus 
we define $f_{R,R^\prime}=1$ (with $R<R^\prime$) if the bond is oriented 
from $R$ to $R^\prime$, and $f_{R,R^\prime}=-1$ otherwise. In summary, in 
order to define the fermionic pairing function $f_{R,R^\prime}$ once we 
know the oriented planar graph, it is necessary to:

\begin{itemize}
\item label the sites according to their lexicographical order,
\item orient the bonds in order to meet the Kasteleyn prescription, and
\item take $f_{R,R^\prime}=1$ for the bond oriented from $R$ to $R^\prime$,
and $f_{R,R^\prime}=-1$ otherwise. 
\end{itemize}

This construction is strictly valid only for planar graphs, namely for graphs 
without intersecting singlets, implying that open boundary conditions must be 
taken. In this case it is known that a unique short-range RVB state can
be constructed. Periodic boundary conditions imply the existence of four 
degenerate states, which are obtained by inserting a cut (changing
the sign of the pairing function on all bonds intersected) that wraps once 
around the system, in the $x$, $y$ or both directions~\cite{chakraborty}. 
These different states have the same bulk properties and, despite the fact 
that it would be possible to obtain a precise correspondence between bosonic 
and fermionic states, their physical properties can be obtained by 
considering a single (bosonic or fermionic) wave function.

\section{Antiferromagnetic order}\label{sec:AF}

In the preceding sections we have considered the mean-field 
Hamiltonian~(\ref{hbcs}) containing only hopping and pairing terms. In 
this case, even by considering the local SU(2) symmetries described above, 
it is not possible to generate a magnetic order parameter. The most natural 
way to introduce an antiferromagnetic order is by adding to the BCS 
Hamiltonian of Eq.~(\ref{hbcs}) a magnetic field
\begin{equation}\label{hbcsaf}
{\cal H}_{BCS+AF} = {\cal H}_{BCS} + {\cal H}_{AF}.
\end{equation}
Usually, the antiferromagnetic mean-field order parameter is chosen is chosen 
to lie along the $z$-direction~\cite{gros},
\begin{equation}\label{deltaz}
{\cal H}_{AF} = \Delta_{AF} \sum_R e^{iQ \cdot R}
(c_{R,\uparrow}^{\dagger}c_{R,\uparrow} -
c_{R,\downarrow}^{\dagger}c_{R,\downarrow}),
\end{equation}
where $Q$ is the antiferromagnetic wave vector (e.g., $Q=(\pi,\pi)$
for the N\'eel state). However, in this case, the Gutzwiller-projected wave 
function obtained from the ground state of Eq.~(\ref{hbcsaf}) overestimates 
the correct magnetic order parameter (see section~\ref{sec:twod}), because 
important quantum fluctuations are neglected. A more appropriate description 
which serves to mitigate this problem is obtained by the introduction of 
a spin Jastrow factor ${\cal J}$ which generates fluctuations in the 
direction \textit{orthogonal} to that of the mean-field order 
parameter~\cite{manousakis,franjio}. Therefore, we take a staggered magnetic 
field $\Delta_{AF}$ along the $x$ axis, 
\begin{equation}\label{deltaxy}
{\cal H}_{AF} = \Delta_{AF} \sum_R e^{iQ \cdot R}
(c_{R,\uparrow}^{\dagger}c_{R,\downarrow} +
c_{R,\downarrow}^{\dagger}c_{R,\uparrow}),
\end{equation}
and consider a long-range spin Jastrow factor ${\cal J}$
\begin{equation}\label{spinjastrow}
{\cal J} = \exp \left ( \frac{1}{2} \sum_{R,R^\prime} v_{R-R^\prime} 
S^z_R S^z_{R^\prime} \right ),
\end{equation}
$v_{R-R^\prime}$ being variational parameters to be optimized by 
minimizing the energy. The Jastrow term is very simple to compute by 
employing a random walk in the configuration space $|x \rangle = 
c_{R_1,\sigma_1}^{\dagger} \dots c_{R_N,\sigma_N}^{\dagger} |0 \rangle$ 
defined by the electron positions and their spin components along the 
$z$ quantization axis, because it represents only a \textit{classical} 
weight acting on the configuration. Finally, the variational \textit{ansatz} 
is given by
\begin{equation}\label{stateaf}
|pBCS+AF \rangle = {\cal J} P_{S_z=0} P_G |BCS+AF \rangle,
\end{equation}
where $P_{S_z=0}$ is the projector onto the $S_z=0$ sector and 
$|BCS+AF \rangle$ is the ground state of the Hamiltonian~(\ref{hbcsaf}).
It should be emphasized that this wave function breaks the spin symmetry,
and thus, like a magnetically ordered state, it is not a singlet. Nevertheless,
after projection onto the subspace with $S^z_{tot}=0$, the wave function has 
$\langle S^x_R\rangle =\langle S^y_R\rangle = \langle S^z_R\rangle =0$.
Furthermore, the correlation functions $\langle S^x_R S^x_{R^\prime}\rangle$ 
and $\langle S^y_R S^y_{R^\prime}\rangle$ have the same behavior, and hence 
the staggered magnetization lies in the $x{-}y$ plane~\cite{manousakis}.

The mean-field Hamiltonian~(\ref{hbcsaf}) is quadratic in the fermionic
operators and can be diagonalized readily in real space. Its ground state 
has the general form
\begin{equation} \label{psimf}
|BCS+AF \rangle = \exp  \left ( \frac{1}{2} 
\sum_{R,R^\prime,\sigma,\sigma^\prime} f^{\sigma,\sigma^\prime}_{R,R^\prime} 
c_{R,\sigma}^{\dagger} c_{R^\prime,\sigma^\prime}^{\dagger} \right ) 
|0 \rangle,
\end{equation}
where the pairing function $f^{\sigma,\sigma^\prime}_{R,R^\prime}$ is an 
antisymmetric $2N \times 2N$ matrix. We note that in the case of the standard 
BCS Hamiltonian, with $\Delta_{AF}=0$ or even with $\Delta_{AF}$ along $z$, 
$f^{\uparrow,\uparrow}_{R,R^\prime}=f^{\downarrow,\downarrow}_{R,R^\prime}=0$,
whereas in the presence of a magnetic field in the $x{-}y$ plane the pairing 
function acquires non-zero contributions also in this triplet channel.
The technical difficulty when dealing with such a state is that, given a
generic configuration with a definite $z$-component of the spin,
$|x \rangle = c_{R_1,\sigma_1}^{\dagger} \dots c_{R_N,\sigma_N}^{\dagger}
|0 \rangle$, one has
\begin{equation} \label{Pfaff}
\langle x|BCS+AF \rangle = Pf[{\bf F}],
\end{equation}
where $Pf[{\bf F}]$ is the Pfaffian of the pairing function
\begin{equation}
{\bf F} \,=\,
\left ( 
\begin{array}{cc} 
\Big [ f(\uparrow,R_\alpha;\uparrow,R_\beta) \Big ] &
\Big [ f(\uparrow,R_\alpha;\downarrow,R^\prime_\beta)\Big ]\cr
\Big [ f(\downarrow,R^\prime_\alpha;\uparrow,R_\beta)\Big ] &
\Big [f(\downarrow,R^\prime_\alpha;\downarrow,R^\prime_\beta)\Big ]
\end{array}\right ),
\end{equation}
in which the matrix ${\bf F}$ has been written in terms of $N\times N$ blocks
and $R_\alpha$ and $R^\prime_\alpha$ are respectively the positions of the up 
and down spins in the configuration $|x \rangle$~\cite{bouchaud}. 

\section{Numerical Results}\label{sec:results}

In this section, we report numerical results obtained by the variational
Monte Carlo method for the one- and two-dimensional lattices. The 
variational parameters contained in the BCS and BCS+AF mean-field 
Hamiltonians of Eqs.~(\ref{hbcs}) and~(\ref{hbcsaf}), as well as the ones 
contained in the spin Jastrow factor,~(\ref{spinjastrow}), can be obtained 
by the optimization technique described in Ref.~\cite{sorella}.

\subsection{One-dimensional lattice}\label{sec:oned}

\begin{figure}[t]
\centering
\includegraphics*[width=.45\textwidth]{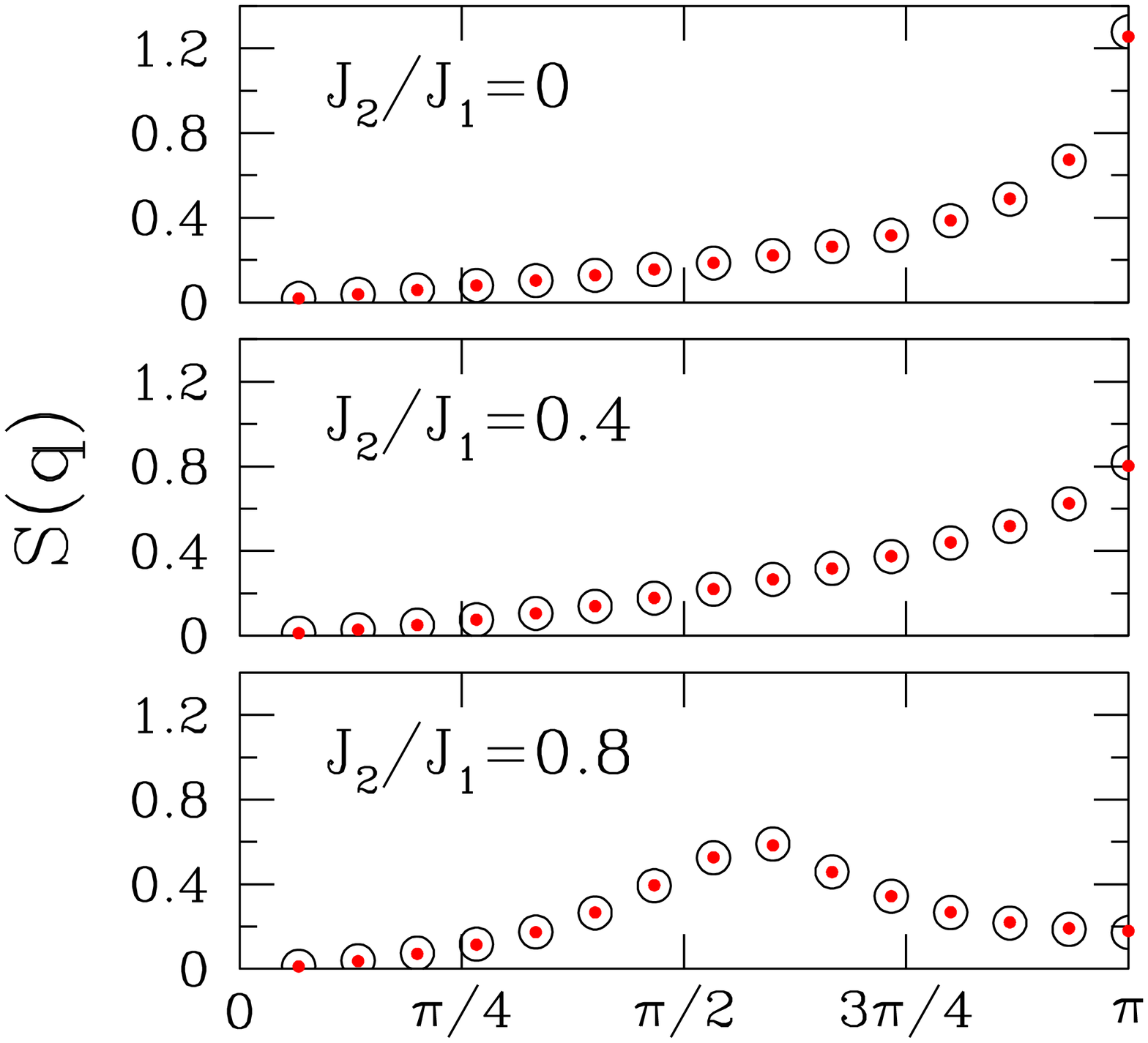}
\includegraphics*[width=.45\textwidth]{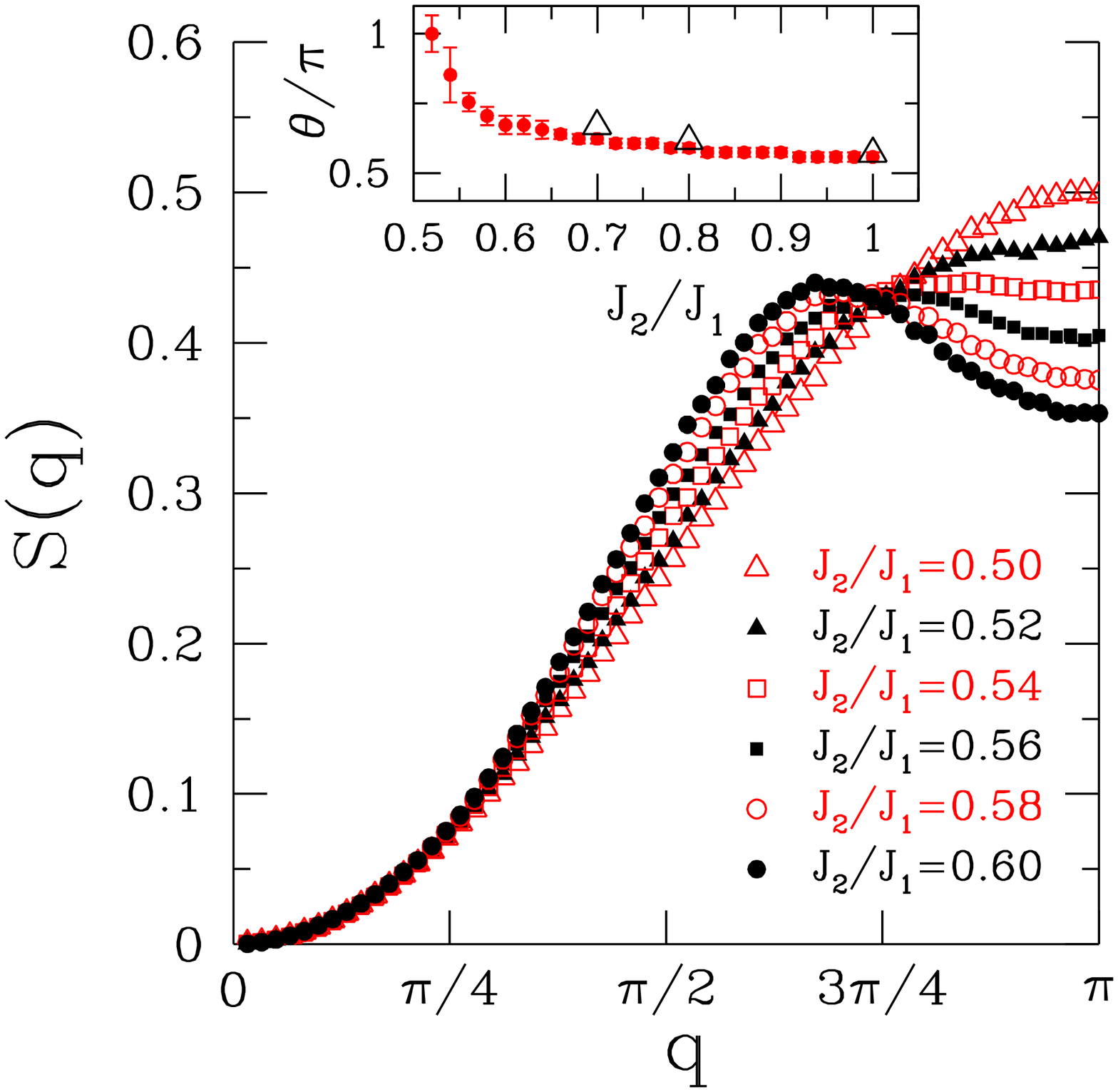}
\caption[]{Left panels: comparison between exact (empty circles) 
and variational (full dots) results for the spin structure factor 
$S(q)$ on a chain with $30$ sites. Right panel: variational results 
for the spin structure factor $S(q)$ for $122$ sites and 
$J_2/J_1 \ge 0.5$. Inset: position of the maximum of $S(q)$,
indicated by $\theta$, as a function of the ratio $J_2/J_1$ (full dots). 
For comparison, the DMRG results of Ref.~\cite{white} are also shown
(empty triangles).}
\label{fig:1-2}
\end{figure}

We begin by considering the one-dimensional case, where the high level of 
accuracy of the pBCS wave function can be verified by comparison with 
Lanczos and DMRG results. We consider the Hamiltonian~(\ref{model}) on a 
chain with $N$ sites and periodic boundary conditions, and first discuss 
in some detail the parametrization of the wave function. For $J_2=0$, a very 
good variational state is obtained simply by projecting the free-electron 
Slater determinant, where $\epsilon_k= -2t \cos k$~\cite{gros2}. Then, in 
one dimension, the nearest-neighbor BCS pairing $\Delta_1$ is irrelevant, 
and, in order to improve the variational energy, a third-neighbor BCS pairing 
$\Delta_3$ must be considered in addition; a second-neighbor pairing term 
$\Delta_2$, like the chemical potential, violates the Marshall-Peierls sign 
rule, which must hold at $J_2=0$, and thus is not considered. To give some 
indication of the accuracy of the wave function, we note that for $N=30$, 
the energy per site of the projected Fermi sea is $E/J_1=-0.443060(5)$, 
while by optimizing $\Delta_1$ and $\Delta_3$ one obtains 
$E/J_1=-0.443934(5)$, the exact result being $E_0/J_1=-0.444065$.

When both the chemical potential and $\Delta_2$ vanish, the particle-hole 
transformation $P_h$ (section~\ref{sec:symm2d}) is a symmetry of the BCS 
Hamiltonian. In finite chains, the BCS ground state is unique only 
if the appropriate boundary conditions are adopted in ${\cal H}_{BCS}$: 
if, for example, $N=4l+2$ with integer $l$, periodic boundary conditions 
(PBC) should be used, while the imposition of antiperiodic boundary 
conditions (APBC) causes four zero-energy modes to appear in the 
single-particle spectrum. By filling these energy levels, we can 
form six orthogonal BCS ground states in the $S_z=0$ subspace, which, in 
the thermodynamic limit, are degenerate with the ground state of the BCS 
Hamiltonian with PBC. However, two of these states have the wrong 
particle-hole quantum number and are therefore annihilated by the 
Gutzwiller projector. If the remaining four BCS states (three singlets 
and one triplet) are still orthogonal after projection, one may infer 
either the presence of a gapless excitation spectrum or of a ground-state 
degeneracy. We have built these five projected states (one with PBC and 
four with APBC) for a $N=30$ chain and variational parameters appropriate 
for $J_2=0$. Two of them belong to the symmetry subspace of the ground 
state and represent the same physical wave function (their overlap 
is $|\langle \Psi_1|\Psi_2 \rangle|=0.999$), two of them are 
singlets with momentum $\pi$ relative to ground state and again show an 
extremely large overlap ($|\langle \Psi_3|\Psi_4 \rangle|=0.921$), and 
the remaining state is a triplet with momentum $\pi$ relative to the 
ground state. Therefore only three independent states can be obtained by 
this procedure. It is remarkable that by optimizing the parameters for the 
ground state, and without any additional adjustable parameters, these three 
variational states have overlap higher than $98.7\%$ with the exact 
eigenstates of the Heisenberg Hamiltonian in the lowest levels of the 
conformal tower of states~\cite{affleck2}, thereby reproducing with high 
accuracy the ground state and the lowest singlet and triplet modes. 

By increasing the frustrating interaction, the parameters $\Delta_1$ and 
$\Delta_3$ (both real) grow until a divergence at $J_2/J_1 \sim 0.15$. For 
larger values of $J_2/J_1$, the band structure changes: here $\epsilon_k
 = -2t^\prime \cos 2k - \mu$ and a non-vanishing BCS pairing is found, 
leading to a finite gap in the BCS spectrum, $E_k=\sqrt{\epsilon_k^2
 + \Delta_k^2}$. In this regime, although the variational wave function is 
translationally invariant, it shows a long-range order in the dimer-dimer 
correlations (see below). Similar behavior has been also discussed in 
Ref.~\cite{sorella2} for a complex wave function on ladders with an odd 
number of legs. The variational parameters appropriate for this regime 
correspond to a gapped BCS single-particle spectrum for both PBC and APBC, 
and then only two states can be constructed. However, the symmetry subspace 
of the variational wave function depends on the choice of boundary conditions, 
implying a ground-state degeneracy. In a chain of $N=30$ sites and for 
$J_2/J_1=0.4$, we found that the two singlets which collapse in the 
thermodynamic limit (due to the broken translational symmetry) have 
overlaps higher than $99\%$ with the two pBCS wave functions corresponding 
to the same variational parameters and different boundary conditions. This 
shows that the pBCS class of wave functions is able to describe valence-bond 
crystals and broken-symmetry states. By increasing further 
the ratio $J_2/J_1$, beyond $0.5$ we found that, while the bare dispersion
is again $\epsilon_k = -2t \cos k$, $\Delta_2$ acquires a finite value 
(together with $\Delta_1$ and $\Delta_3$), showing both dimerization 
and short-range incommensurate spin correlations.

The primary drawback of the variational scenario is that the critical point
for the transition from the gapless fluid to the dimerized state is predicted 
around $J_2/J_1 \sim 0.15$ (where the best singlet variational state is the 
fully projected Fermi sea), considerably smaller than the known critical 
point $J_2/J_1 \sim 0.241$. This estimate does not change appreciably on 
considering further parameters in the BCS Hamiltonian~(\ref{hbcs}), probably 
because the variational wave function does not describe adequately the 
backscattering term which is responsible for the transition~\cite{parola}. 
In order to improve this aspect, it is necessary to 
include the spin Jastrow factor of Eq.~(\ref{spinjastrow}) (without the
mean-field magnetic parameter $\Delta_{AF}$). In this way, although the 
variational state is no longer a singlet, the value of the square of the 
total spin $\langle S^2\rangle$ remains very small (less than $0.002$ and 
$0.02$ for $30$ and $122$ sites, respectively) and no long-range magnetic 
order is generated. The Jastrow factor is particularly important in the 
gapless regime: despite the fact that the gain in energy with respect to 
the singlet state is less than $10^{-4}J_1$ (specifically, for $J_2=0$ we 
obtain $E/J_1=-0.444010(5)$), this correction is able to shift the 
transition, always marked by the divergence of the BCS pairings, to 
$J_2/J_1 \sim 0.21$, a value much closer to the exact result. A finite 
value of the chemical potential is generated for $0.22 < J_2/J_1 < 0.5$.

\begin{figure}[t]
\centering
\includegraphics*[width=.45\textwidth]{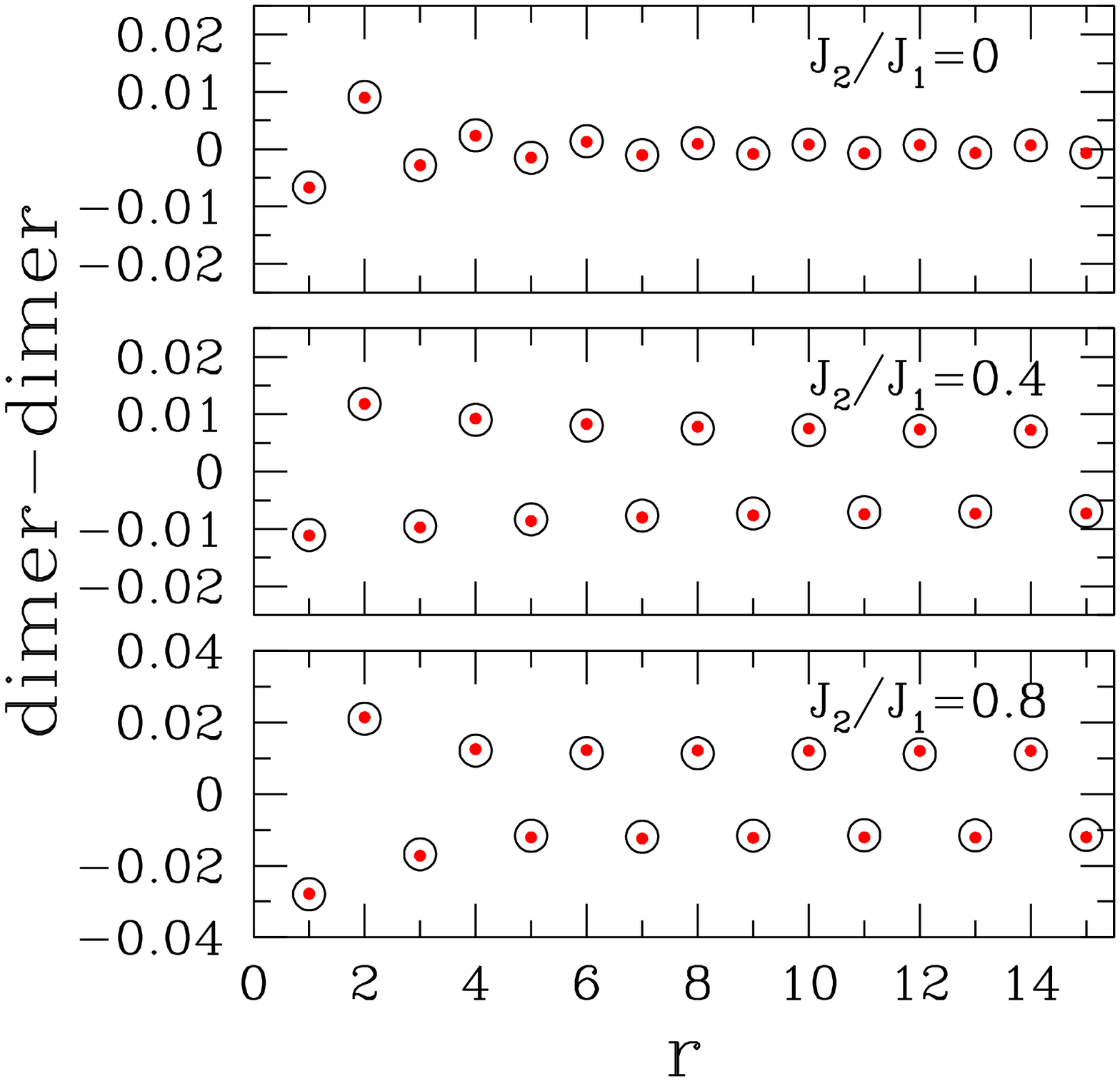}
\includegraphics*[width=.45\textwidth]{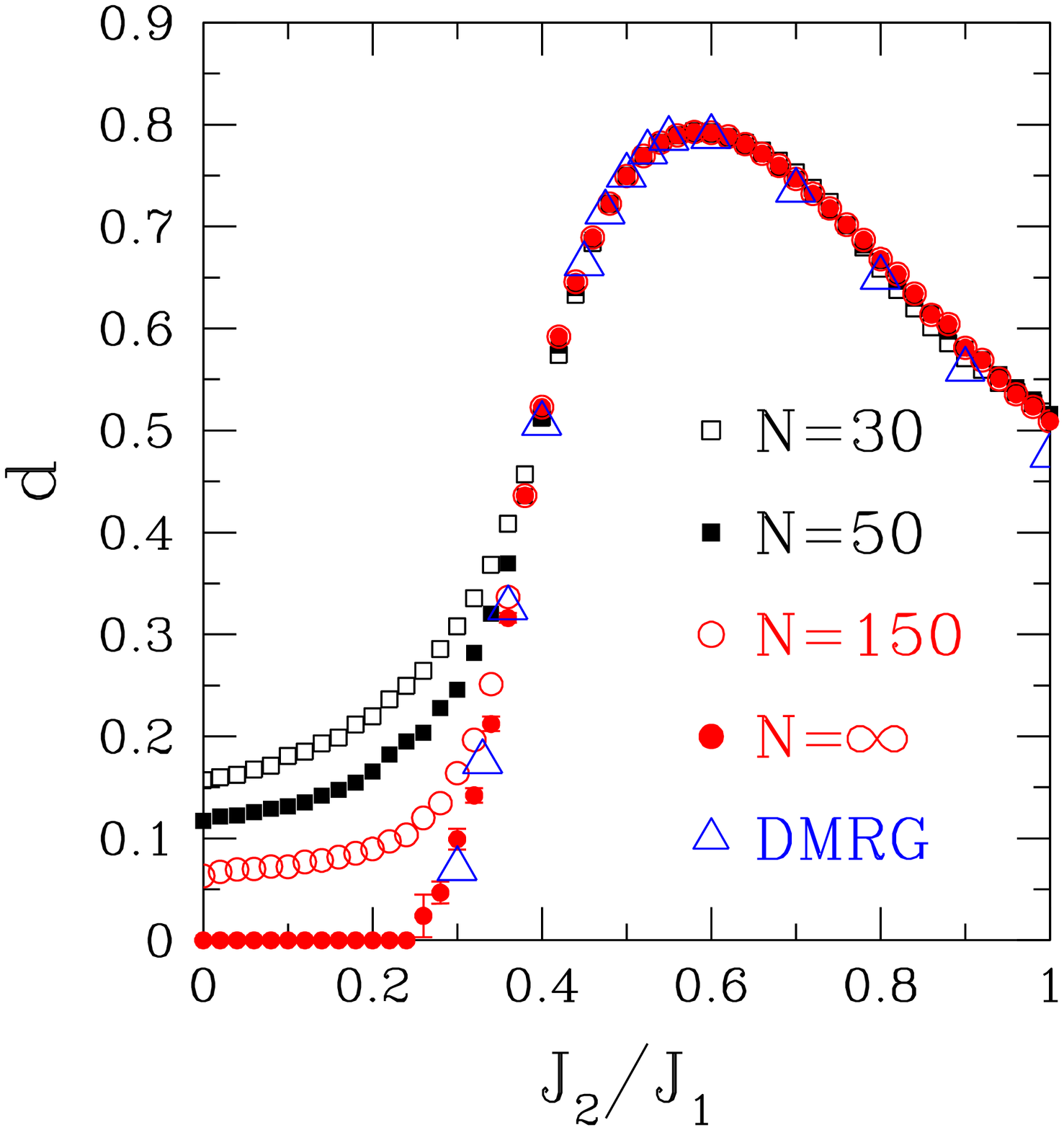}
\caption[]{Left panels: dimer-dimer correlations as a function of 
distance for exact (empty circles) and variational (full dots) 
calculations on a chain of $30$ sites. Right panel: dimer order parameter 
of Eq.~(\ref{dimord}) as a function of the ratio $J_2/J_1$ for $N=30$, 
$50$, and $150$; the extrapolation to the $N\to \infty$ limit is also 
shown, together with the DMRG results of Ref.~\cite{white}.}
\label{fig:3-4}
\end{figure}

Let us now investigate the physical properties of the variational wave 
function by evaluating some relevant correlation functions. The spin 
structure factor is defined as
\begin{equation}
S(q) = \frac{1}{N}\sum_{R,R^\prime} e^{iq(R-R^\prime)} \langle S_R^z 
S_{R^\prime}^z \rangle.
\end{equation}
While true long-range magnetic order cannot be established in 
one-dimensional systems, for $J_2/J_1 \ll 1$ the ground state is 
quasi-ordered, by which is meant that it sustains zero-energy excitations 
and $S(q)$ displays a logarithmic divergence at $q=\pi$. 
In Fig.~\ref{fig:1-2}, we show the comparison of the spin structure factor 
for an exact calculation on $N=30$ and for the variational wave function.
Remarkably, the variational results deliver a very good description of 
$S(q)$ in all the different regimes: for small $J_2/J_1$, where the spin 
fluctuations are commensurate and there is a quasi-long-range order, for 
$0.21 < J_2/J_1 < 0.5$, where the spin fluctuations are still commensurate 
but short-range, and for $J_2/J_1 > 0.5$, where they are incommensurate and 
the maximum of $S(q)$ moves from $q=\pi$ at $J_2/J_1=0.5$ to $q=\pi/2$ for 
$J_2/J_1 \to \infty$. Indeed, it is known that the quantum case is rather 
different from its classical counterpart~\cite{white}: while the latter
shows a spiral state for $J_2/J_1 > 0.25$, with a pitch angle $\theta$ 
given by $\cos \theta = - J_1/4J_2$, the former maintains commensurate 
fluctuations at least up to the Majumdar-Ghosh point. The behavior of 
$S(q)$ for a large lattice with $122$ sites and $J_2/J_1 > 0.5$ is shown 
in Fig.~\ref{fig:1-2}, where we find good agreement with previous 
numerical results based upon the DMRG technique~\cite{white}. 

In the one-dimensional $J_1{-}J_2$ model, there is clear evidence
for a Berezinskii-Kosterlitz-Thouless transition on increasing the ratio
$J_2/J_1$ from a gapless Luttinger liquid to a dimerized state that breaks
the translational symmetry. In order to investigate the possible occurrence 
of a dimerized phase, we analyze the dimer-dimer correlation functions of the 
ground state,
\begin{equation}\label{dimdim}
\Theta(R-R^\prime) 
=\langle S_R^z S_{R+x}^z S_{R^\prime}^z S_{R^\prime+x}^z\rangle -
\langle S_R^z S_{R+x}^z \rangle \langle S_{R^\prime}^z S_{R^\prime+x}^z\rangle.
\end{equation}
While this definition considers only the $z$ component of the spin operators, 
in the presence of a broken spatial symmetry the transverse components must 
also remain finite at large distances, displaying also a characteristic 
alternation. By contrast, in the gapless regime, the dimer correlations 
decay to zero at large distances. The differing behavior of these 
correlations is easy to recognize, with oscillatory power-law decay in the 
Luttinger regime and constant-amplitude oscillations in the dimerized phase. 
Figure \ref{fig:3-4} illustrates the comparison of the dimer-dimer 
correlations~(\ref{dimdim}) between the exact and the variational results 
on a chain with $30$ sites. Also for this quantity we obtain very good 
agreement for all values of the frustrating superexchange $J_2$, both in 
the gapless and in the dimerized regions. Following Ref.~\cite{white}, 
it is possible by finite-size scaling to obtain an estimate of the dimer 
order parameter from the long-distance behavior of the dimer-dimer 
correlations,
\begin{equation}\label{dimord}
d^2=9 \lim_{|R|\to\infty} \vert (\Theta(R-x)-2\Theta(R)+\Theta(R+x)\vert,
\end{equation}
where the factor $9$ is required to take into account the fact that in
Eq.~(\ref{dimdim}) we considered only the $z$ component of the spin operators.
In Fig.~\ref{fig:3-4}, we present the values of the dimer order parameter as 
a function of $J_2/J_1$ for three different sizes of the chain, and also the 
extrapolation in the thermodynamic limit, where the agreement with the DMRG 
results of Ref.~\cite{white} is remarkable.

\subsection{Two-dimensional lattice}\label{sec:twod}

\begin{figure}[t]
\centering
\includegraphics*[width=.45\textwidth]{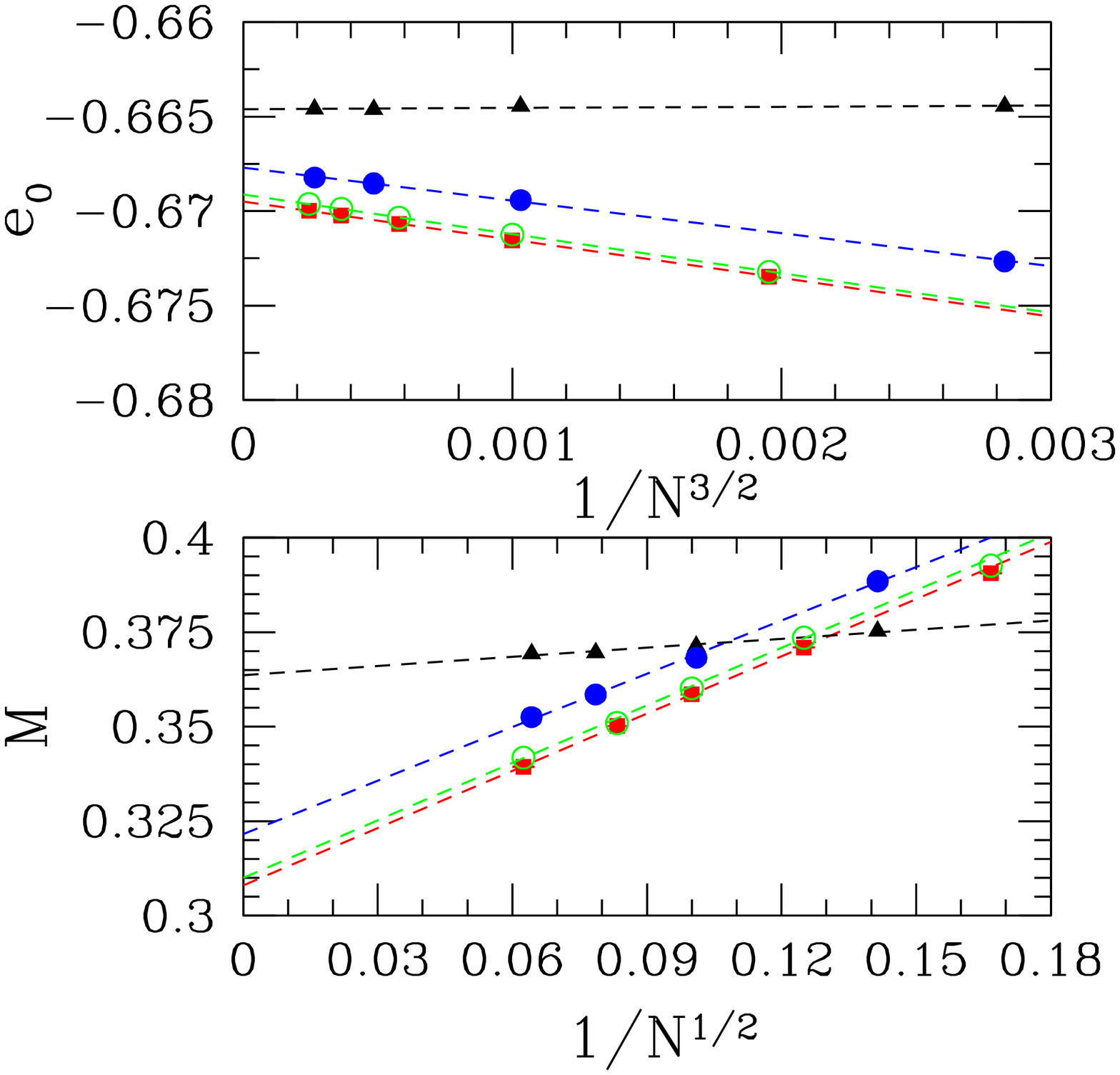}
\includegraphics*[width=.45\textwidth]{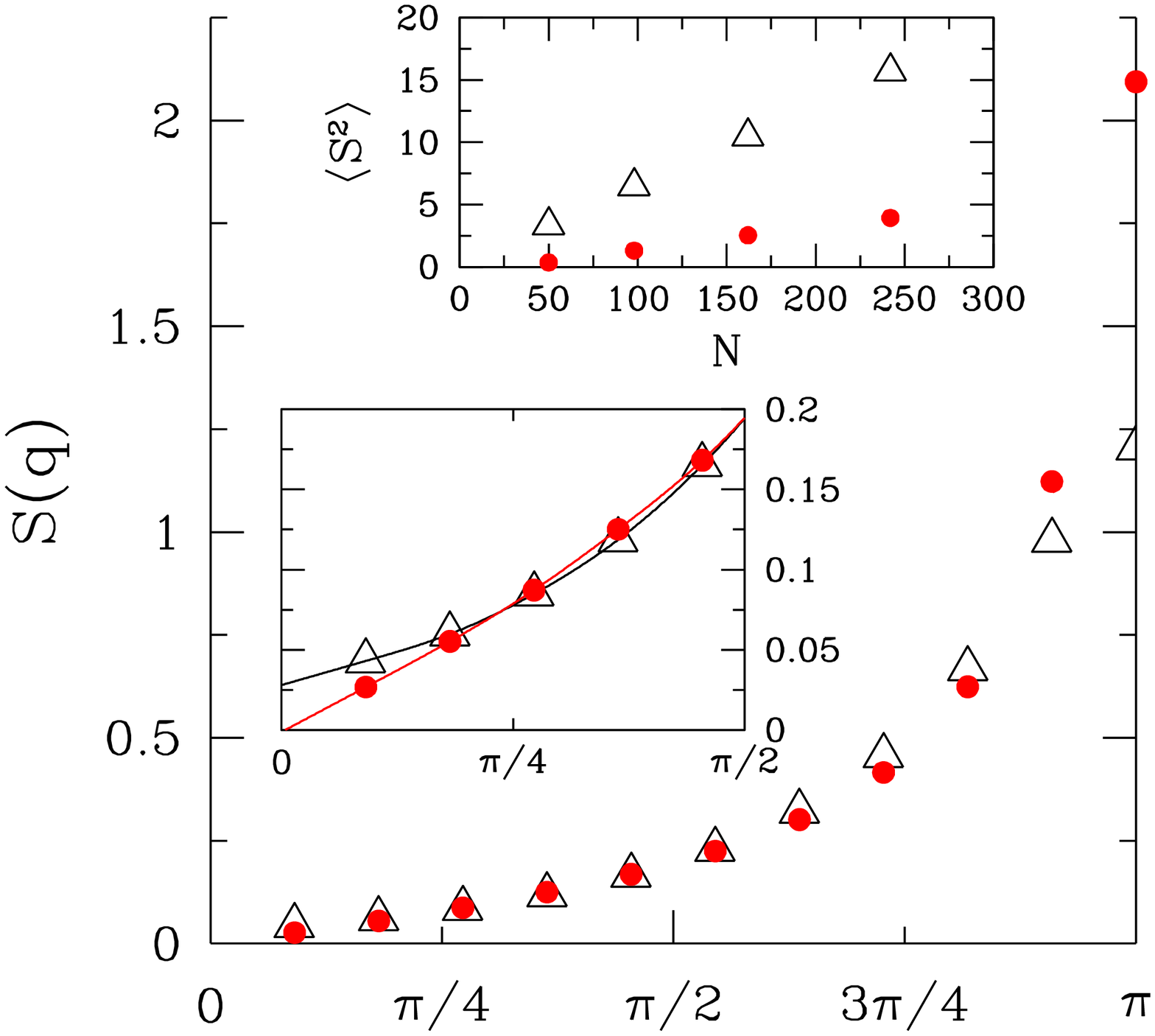}
\caption[]{Upper left panel: energy per site as a function of cluster 
size $N$, showing exact results (full squares), variational results obtained 
by considering Eqs.~(\ref{hbcsaf}) and~(\ref{deltaxy}) with the spin Jastrow 
factor~(\ref{spinjastrow}) (full circles), and variational results obtained 
with Eqs.~(\ref{hbcsaf}) and~(\ref{deltaz}) (full triangles). The results 
obtained by optimizing the pairing function $f_{R,R^\prime}^{bos}$ of the 
bosonic representation, described in section~\ref{sec:boson}, are also 
shown (empty circles)~\cite{sandvik4}. Lower left panel: staggered 
magnetization with the same notation as in the upper panel.

Right panel: static structure factor $S(q)$ for a cluster with 
$N=242$ (tilted by $45^\circ$): variational results for the state of 
Eqs.~(\ref{hbcsaf}) and~(\ref{deltaxy}) with a long-range Jastrow factor 
(full dots) and for the wave function of Eqs.~(\ref{hbcsaf}) and~(\ref{deltaz})
with a nearest-neighbor Jastrow factor (empty triangles).
Lower inset: detail at small momenta. 
Upper inset: square of total spin $\langle S^2 \rangle$ as a function of $N$ 
for the two states, using the same symbols.}
\label{fig:5-6}
\end{figure}

We move now to consider the two-dimensional case, starting with 
the unfrustrated model ($J_2=0$), for which exact results can be 
obtained by Monte Carlo methods~\cite{reger,sandvik3,calandra}. In the 
thermodynamic limit, the ground state is antiferromagnetically ordered with 
a staggered magnetization reduced to approximately $60\%$ of its classical 
value, namely $M \simeq 0.307$~\cite{sandvik3,calandra}. This quantity can 
be obtained both from the spin-spin correlations at the largest distances and 
from the spin structure factor $S(q)$ at $q=(\pi,\pi)$. In the following, we 
will consider the former definition and will calculate the isotropic 
correlations $\langle {\bf S}_R \cdot {\bf S}_{R^\prime} \rangle$, because 
this quantity is known to have smaller finite-size effects~\cite{reger, 
sandvik3}. For the unfrustrated case, the best wave function has $\epsilon_k
 = -2t(\cos k_x+\cos k_y)$ and a pairing function with $d_{x^2-y^2}$ symmetry, 
$\Delta^{x^2-y^2}_k = \Delta_1 (\cos k_x-\cos k_y)$ (possibly also with higher 
harmonics connecting opposite sublattices). The quantity $\Delta_{AF}$ in 
Eq.~(\ref{deltaxy}) has a finite value and the spin Jastrow 
factor~(\ref{spinjastrow}) has an important role.

Figure~\ref{fig:5-6} shows the comparison of the variational calculations
with the exact results, which are available for rather large system sizes.
In the unfrustrated case, the bosonic representation is considerably better 
than the fermionic one: the accuracy in the energy is around $0.06\%$ and 
the sublattice magnetization is also very close to the exact 
value~\cite{lou,sandvik4}. However, the fermionic state defined 
by Eqs.~(\ref{hbcsaf}) and~(\ref{deltaxy}), in combination with the spin 
Jastrow factor, also provides a very good approximation to the exact 
results (energy per site and staggered magnetization), whereas the 
wave function defined by Eqs.~(\ref{hbcsaf}) and~(\ref{deltaz}) is rather 
inaccurate. It should be emphasized that when the Jastrow factor is included, 
the slopes of the finite-size scaling functions are also remarkably similar 
to the exact ones, both for the energy per site $e_0$ and for the 
magnetization $M$. This implies that the pBCS wave function provides an 
accurate estimate of the spin velocity $c$, of the transverse susceptibility 
$\chi_\perp$, and as a consequence of the spin stiffness, $\rho_s=c^2 
\chi_\perp$. By contrast, the wave function without the Jastrow factor 
leads to a vanishing spin velocity.  We note that in this case the 
staggered magnetization $M \simeq 0.365$ is also overestimated in the 
thermodynamic limit.

\begin{figure}[t]
\centering
\includegraphics*[width=.45\textwidth]{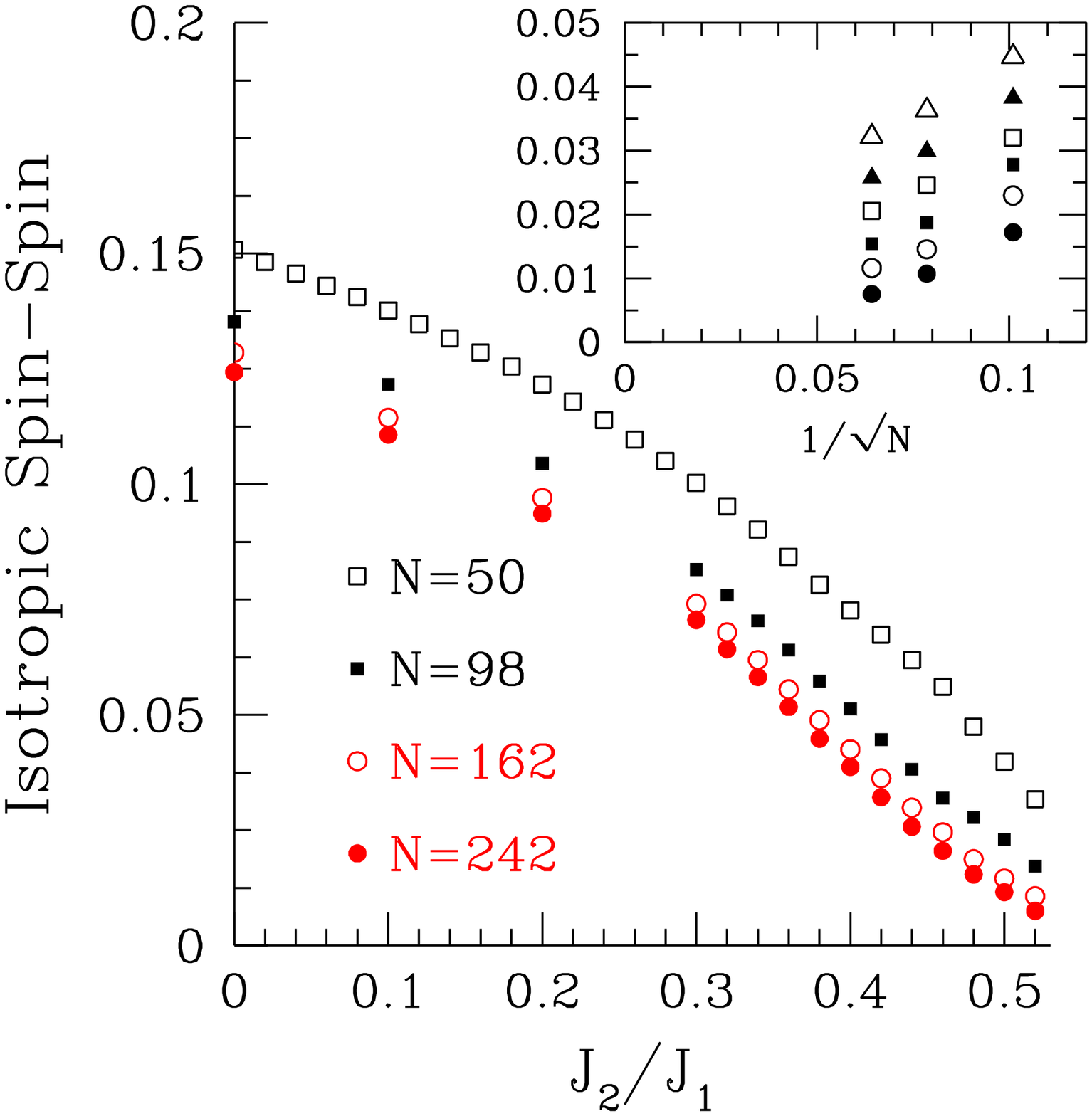}
\includegraphics*[width=.45\textwidth]{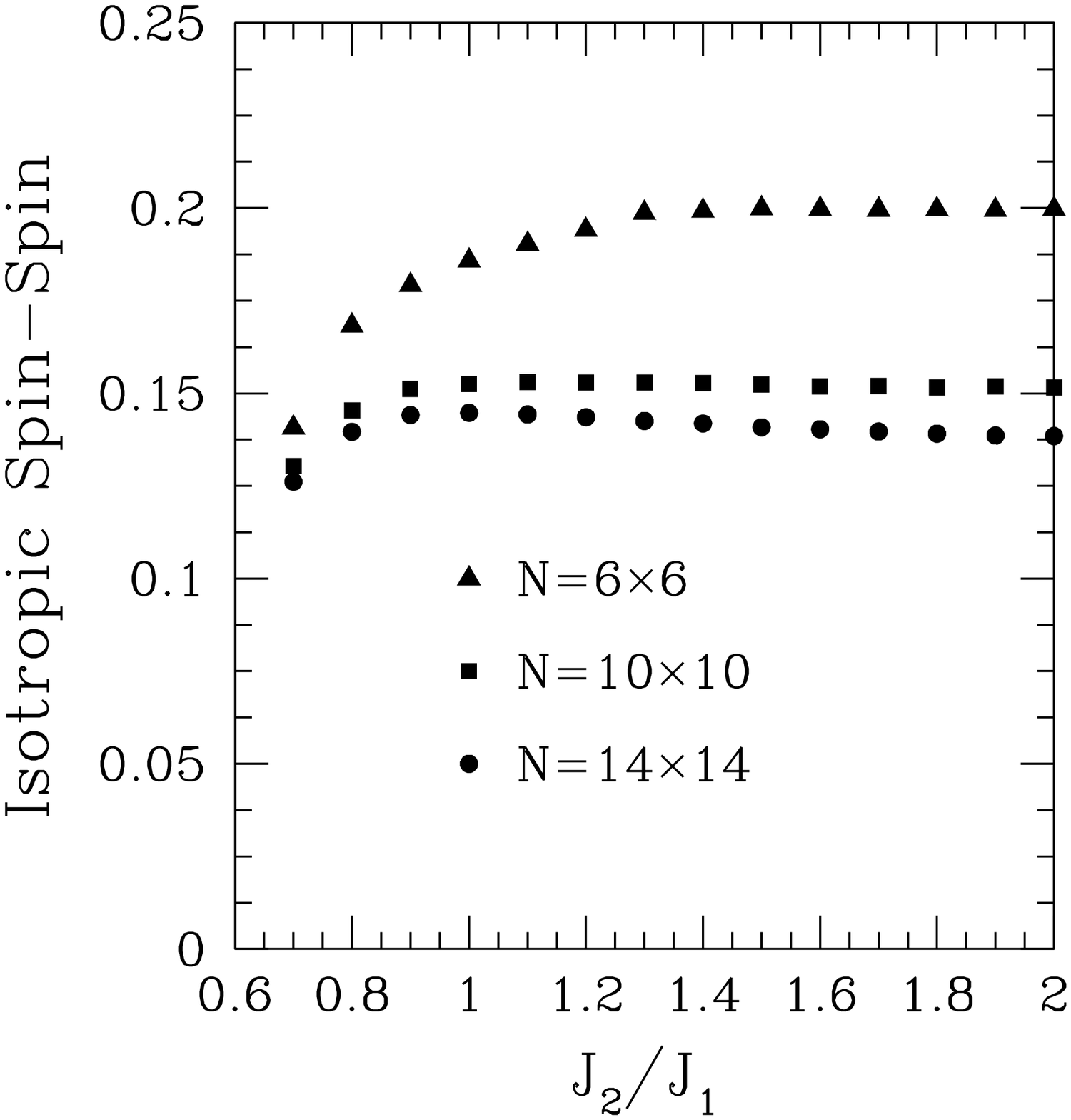}
\caption[]{Left panel: spin-spin correlations at the largest distances
as a function of the ratio $J_2/J_1$ for different cluster sizes $N$. 
Inset: finite-size scaling for $J_2/J_1=0.42$ (empty triangles), $0.44$ 
(full triangles), $0.46$ (empty squares), $0.48$ (full squares), 
$0.50$ (empty circles), and $0.52$ (full circles). Right panel:
spin-spin correlations at the largest distances for $J_2/J_1>0.7$.}
\label{fig:7-8}
\end{figure}

The functional form of the Jastrow factor at long ranges, which can be 
obtained by minimizing the energy, is necessary to reproduce correctly the 
small-$q$ behavior of the spin-structure factor $S(q)$, mimicking the 
Goldstone modes typical of a broken continuous symmetry~\cite{manousakis}. 
Indeed, it is clear from Fig.~\ref{fig:5-6} that only with a long-range spin 
Jastrow factor it is possible to obtain $S(q) \sim |q|$ for small momenta, 
consistent with a gapless spin spectrum. By contrast, with a short-range 
spin Jastrow factor (for example with a nearest-neighbor term), 
$S(q) \sim {\rm const}$ for small $q$, which is clearly not 
correct~\cite{manousakis}. Finally, it should be emphasized that the 
combined effects of the magnetic order parameter $\Delta_{AF}$ and the 
spin Jastrow factor give rise to an almost singlet wave function, strongly 
reducing the value of $\langle S^2\rangle$ compared to the case without a 
long-range Jastrow term (see Fig.~\ref{fig:5-6}).

On increasing the value of the frustrating superexchange $J_2$, the Monte 
Carlo method is no longer numerically exact because of the sign problem, 
whereas the variational approach remains easy to apply. In Fig.~\ref{fig:7-8}, 
we present the results for the spin-spin correlations at the maximum 
accessible distances for $J_2/J_1 \le 0.52$. It is interesting to note that 
when $J_2/J_1>0.4$, a sizable energy gain may be obtained by adding a finite 
pairing connecting pairs on the same sublattice with $d_{xy}$ symmetry, 
namely $\Delta_k = \Delta^{x^2-y^2}_k + \Delta^{xy}_k$~\cite{capriotti}. 
The mean-field order parameter $\Delta_{AF}$ remains finite up to 
$J_2/J_1 \sim 0.5$, whereas for $J_2/J_1 > 0.5$ it goes to zero in the 
thermodynamic limit. Because the Jastrow factor is not expected to destroy 
the long-range magnetic order, the variational technique predicts that 
antiferromagnetism survives up to higher frustration ratios than 
expected~\cite{doucot}, similar to the outcome of a Schwinger boson 
calculation~\cite{mila}. The magnetization also remains finite,
albeit very small, up to $J_2/J_1=0.5$ (see Fig.~\ref{fig:7-8}). We 
remark here that by using the bosonic RVB state, Beach argued that the 
Marshall-Peierls sign rule may hold over a rather large range of frustration,
namely up to $J_2/J_1=0.418$, also implying a finite staggered magnetic
moment~\cite{beach}. In this approach, if one assumes a continuous 
transition from the ordered to the disordered phase, the critical value 
is found to be $J_2/J_1=0.447$, larger than the value of Ref.~\cite{doucot} 
and much closer to our variational prediction. We note in this context that 
recent results obtained by coupled cluster methods are also similar,
i.e., $J_2/J_1 \sim 0.45$ for a continuous phase transition between a 
N\'eel ordered state and a quantum paramagnet~\cite{darradi}.

\begin{figure}[t]
\centering
\includegraphics*[width=.5\textwidth]{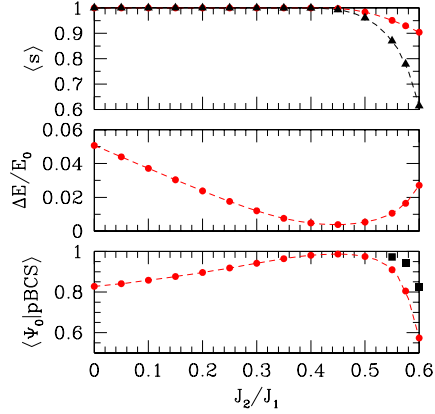}
\caption[]{Comparison between exact and variational results on a
$6 \times 6$ lattice. The pBCS wave function has $\Delta_{AF}=0$ and 
no Jastrow factor. Upper panel: average sign $\langle s \rangle$ of
Eq.~(\ref{signo}) (full circles); the Marshall-Peierls sign,
$\langle s \rangle_{MA} = \sum_x |\langle x|\Psi_0\rangle|^2 
{\rm sign} \left\{ \langle x|\Psi_0\rangle (-1)^{N_\uparrow(x)}\right\}$, 
is also shown (full triangles). Middle panel: accuracy of the ground-state 
energy, $\Delta E/E_0=(E_0-E_{pBCS})/E_0$, where $E_0$ and $E_{pBCS}$ are the 
exact and the variational energies, respectively. Lower panel: overlap between 
the exact $|\Psi_0\rangle$ and variational $|pBCS\rangle$ states 
(full circles). The norm of the projection of the variational state onto the 
subspace spanned by the two lowest-energy states in the same symmetry sector 
is also shown (full squares) close to the first-order transition to the 
collinear state.}
\label{fig:9}
\end{figure}

\begin{figure}[t]
\centering
\includegraphics*[width=.45\textwidth]{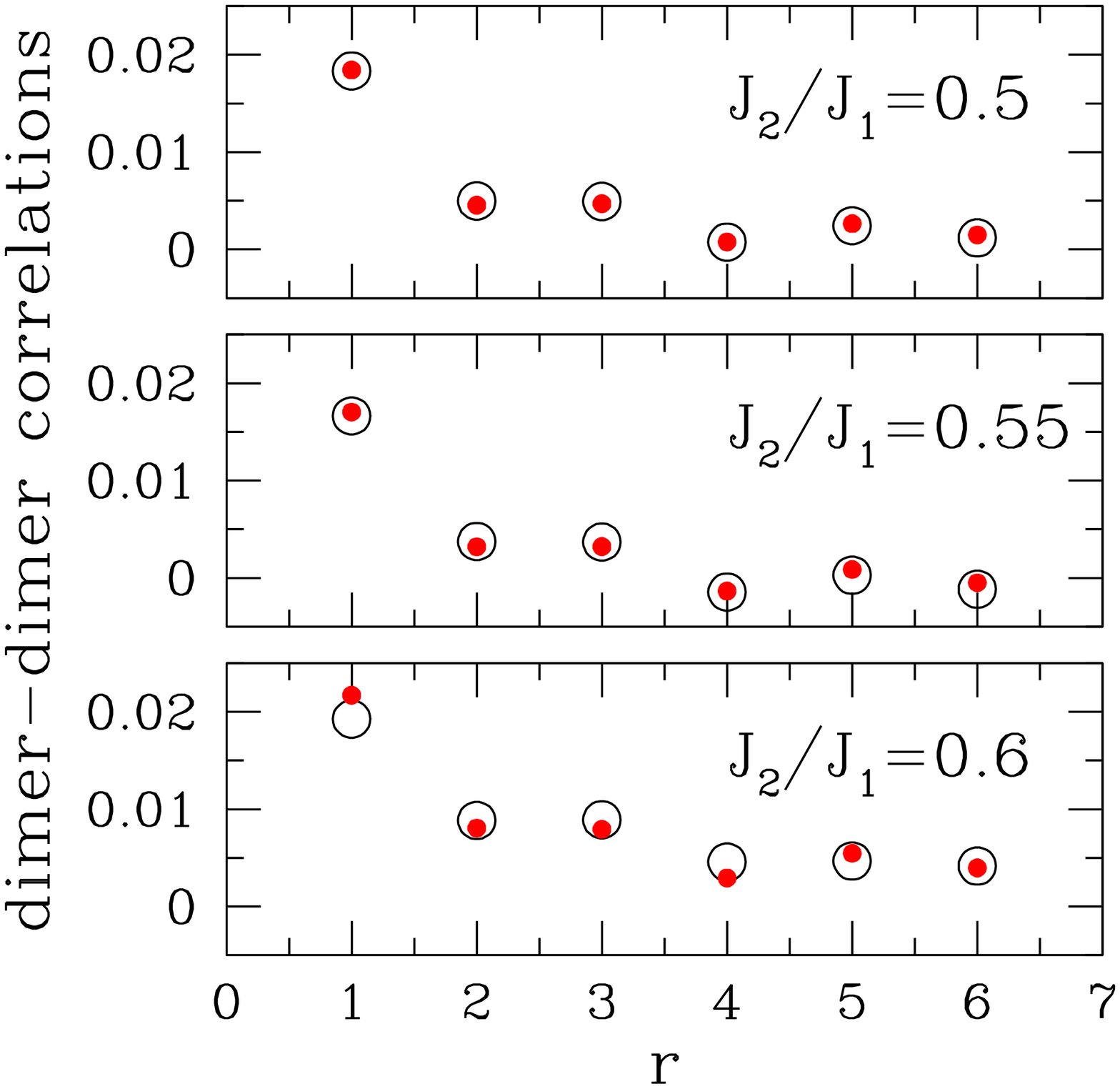}
\includegraphics*[width=.45\textwidth]{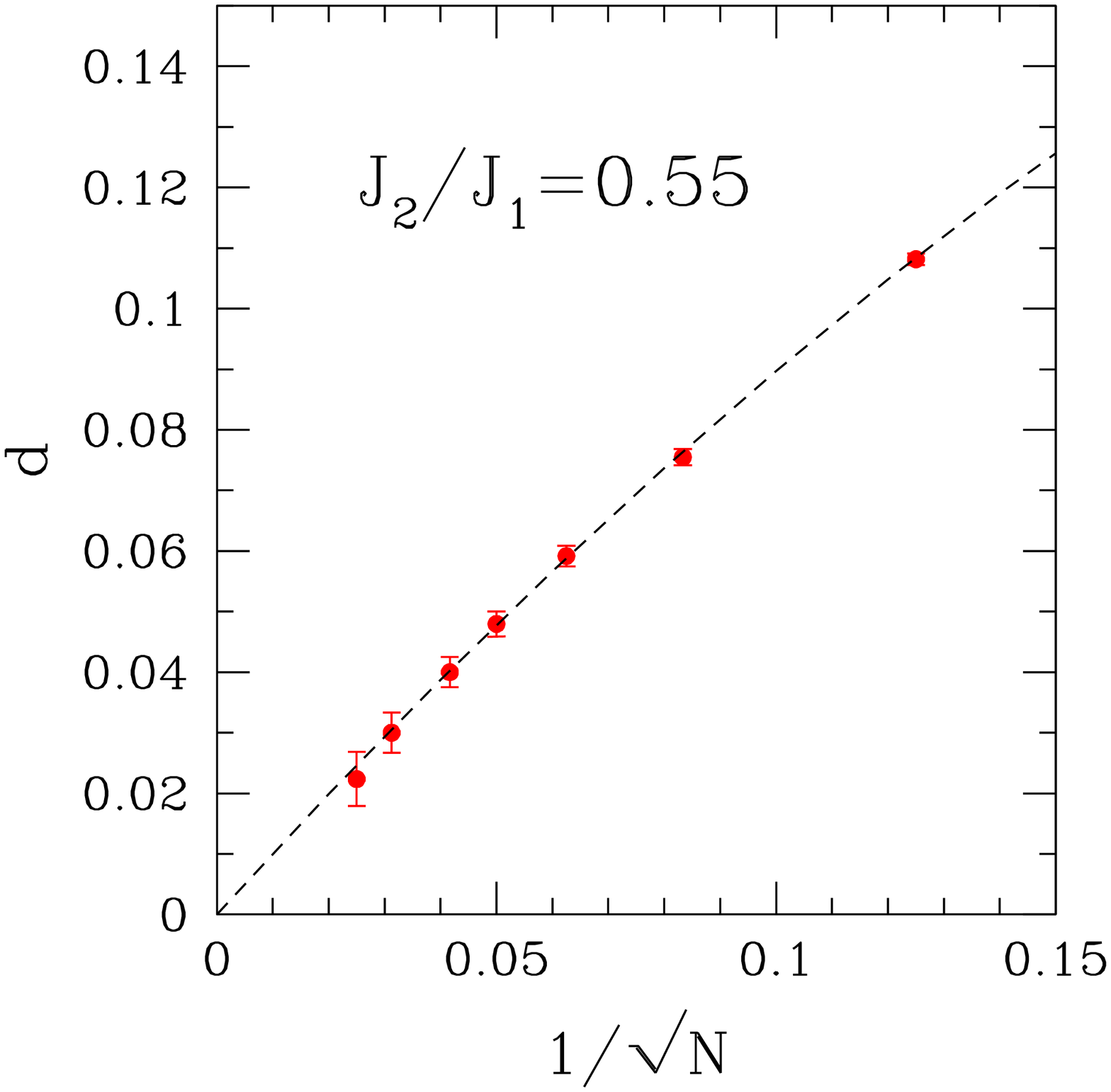}
\caption[]{Left panels: dimer-dimer correlations as a function of the 
Manhattan distance for exact (empty circles) and variational (full dots) 
calculations on a $6 \times 6$ cluster. Right panel: finite-size scaling 
of the dimer order parameter for $J_2/J_1=0.55$.}
\label{fig:10-11}
\end{figure}

\begin{table}[!pb]
\centering
\caption[]{Energies per site for a $6 \times 6$ lattice. $E_{pBCS}$ obtained 
from the pBCS wave function (with $\Delta_{AF}=0$ and no Jastrow factor), 
$E_{LR-RVB}$ from the long-range bosonic RVB state, optimizing just one 
parameter using the master-equation method~\cite{beach2}, and $E_{SR-RVB}$ 
obtained by diagonalizing the $J_1{-}J_2$ model in the nearest-neighbor 
valence-bond basis~\cite{mambrini2}. The exact results $E_0$ are also 
reported.}
\renewcommand{\arraystretch}{1.2}
\setlength\tabcolsep{5pt}
\begin{tabular}{@{}lllll@{}}
\hline\noalign{\smallskip}
$J_2/J_1$ & $E_{SR-RVB}$ & $E_{LR-RVB}$ & $E_{pBCS}$   & $E_0$ \\
\hline\noalign{\smallskip}
0.30      &  -0.54982    & -0.5629(5)   & -0.55569(2) & -0.56246 \\
0.35      &  -0.53615    & -0.5454(5)   & -0.54134(1) & -0.54548 \\
0.40      &  -0.52261    & -0.5289(5)   & -0.52717(1) & -0.52974 \\
0.45      &  -0.50927    &              & -0.51365(1) & -0.51566 \\
0.50      &  -0.49622    &              & -0.50107(1) & -0.50381 \\
0.55      &  -0.48364    &              & -0.48991(1) & -0.49518 \\
0.60      &  -0.47191    &              & -0.47983(2) & -0.49324 \\
\hline
\end{tabular}
\label{tab:1}
\end{table}

In the regime of large $J_2/J_1$ (i.e., $J_2/J_1 > 0.65$), collinear order
with pitch vectors $Q=(0,\pi)$ and $Q=(\pi,0)$ is expected. The pBCS 
wave function is also able to describe this phase through 
a different choice for the bare electron dispersion, namely
$\epsilon_k=-2t^\prime [\cos(k_x+k_y) + \cos(k_x-k_y)]$ and
$\Delta_k = \Delta_1 \cos k_x +\Delta_2 [\cos(k_x+k_y) - \cos(k_x-k_y)]$, 
with $\Delta_1 \to 0$ for $J_1/J_2 \to 0$. Further, the antiferromagnetic 
wave vector $Q$ in Eq.~(\ref{deltaxy}) is $Q=(\pi,0)$. The variational wave 
function breaks the reflection symmetry of the lattice and, in finite systems,
its energy can be lowered by projecting the state onto a subspace of definite 
symmetry. The results for the spin-spin correlations are shown in 
Fig.~\ref{fig:7-8}. By decreasing the value of $J_2/J_1$, we find clear 
evidence of a first-order phase transition, in agreement with previous 
calculations using different approaches~\cite{singh2,sushkov}.

For $0.5 < J_2/J_1 < 0.65$, the best variational wave function has no
magnetic order ($\Delta_{AF}=0$ and no Jastrow factor) and the BCS 
Hamiltonian has $\epsilon_k=-2t (\cos k_x + \cos k_y)$ and $\Delta_k = 
\Delta^{x^2-y^2}_k + \Delta^{xy}_k$, where $\Delta^{x^2-y^2}_k$ connects 
pairs on opposite sublattices while $\Delta^{xy}_k$ is for same sublattice. 
With this specific electron pairing, the signs of the wave function are
different from those predicted by the Marshall-Peierls rule and are much 
more similar to the exact ones. We define
\begin{equation}\label{signo}
\langle s \rangle = \sum_x |\langle x|pBCS \rangle|^2 
{\rm sign} \left\{ \langle x|pBCS \rangle \langle x|\Psi_0\rangle \right\},
\end{equation}
where $|pBCS \rangle$ and $|\Psi_0\rangle$ are the variational and the exact 
states, respectively. This quantity is shown in Fig.~\ref{fig:9}, together 
with the Marshall-Peierls sign, for a $6 \times 6$ lattice. The variational 
energy, the very large overlap with the exact ground state, and the 
dimer-dimer correlations shown in Figs.~\ref{fig:9} and~\ref{fig:10-11}, 
all reflect the extremely high accuracy of this state in the strongly 
frustrated regime. On small clusters, the overlap between the variational 
wave function and the ground state deteriorates for $J_2/J_1 > 0.55$. This 
may be a consequence of the proximity to the first order transition, which 
marks the onset of collinear magnetic order, and implies a mixing of the 
two finite-size ground states corresponding to the coexisting phases. 

\begin{figure}[t]
\centering
\includegraphics*[width=.45\textwidth]{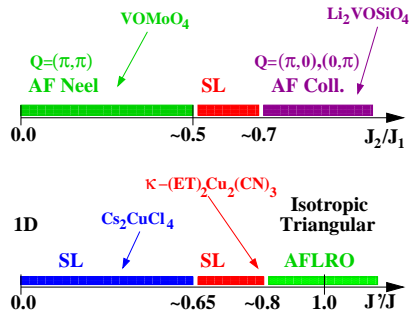}
\caption[]{Upper panel: phase diagram of the $J_1{-}J_2$ model on the
square lattice, as deduced from the variational approach. Lower panel: 
phase diagram of the anisotropic triangular lattice from 
Ref.~\cite{yunoki}. The approximate locations of some relevant materials 
are indicated by the arrows.}
\label{fig:12}
\end{figure}

In Table~\ref{tab:1}, we report the comparison between the energies of the
non-magnetic pBCS wave function and two bosonic RVB states. The first 
is obtained by a full diagonalization of the $J_1{-}J_2$ model in the 
nearest-neighbor valence-bond basis, namely by optimizing all the amplitudes 
of the independent valence-bond configurations without assuming the particular 
factorized form of Eq.~(\ref{bosonicrvb})~\cite{mambrini2}. Although this wave
function contains a very large number of free parameters, its energy is
always higher than that obtained from the pBCS state, showing the
importance of having long-range valence bonds. A further drawback of this 
approach is that it is not possible to perform calculations on large system 
sizes, the upper limit being $N \sim 40$. The second RVB state is obtained 
by considering long-range valence bonds, with their amplitudes given by 
Eq.~(\ref{bosonicrvb}) and optimized by using the master-equation 
scheme~\cite{beach2}. While this wave function is almost exact in the 
weakly frustrated regime, its accuracy deteriorates on raising the 
frustrating interaction, and for $J_2/J_1 > 0.425$ the minus-sign problem 
precludes the possibility of reliable results.
On the other hand, the pBCS state (without antiferromagnetic order or 
the Jastrow term) becomes more and more accurate on approaching the 
disordered region. Remarkably, for $J_2/J_1=0.4$, the energy per site in the 
thermodynamic limit obtained with the long-range bosonic wave function is
$E/J_1=-0.5208(2)$, which is very close to and only slightly higher than 
that obtained from the fermionic representation, $E/J_1=-0.5219(1)$.

In the disordered phase, the pBCS wave function does not break any lattice 
symmetries (section~\ref{sec:symm2d}) and does not show any tendency towards
a dimerization. Indeed, the dimer order parameter $d$ (calculated from the 
correlations at the longest distances) vanishes in the thermodynamic limit, 
as shown in Fig.~\ref{fig:10-11}, implying a true spin-liquid phase in this 
regime of frustration. This fact is in agreement with DMRG calculations on 
ladders with odd numbers of legs, suggesting a vanishing spin gap for all 
values of $J_2/J_1$~\cite{capriotti2}, in sharp contrast to the dimerized 
phase, which has a finite triplet gap. 

Taking together all of the above results, it is possible to draw the 
(zero-temperature) phase diagram generated by the variational approach, 
and this is shown in Fig.~\ref{fig:12}.

We conclude by considering the important issue of the low-energy spectrum. 
In two dimensions, it has been argued that the ground state of a spin-1/2
system is either degenerate or it sustains gapless excitations~\cite{hastings},
in analogy to the one-dimensional case~\cite{lsm}. In Ref.~\cite{ivanov},
it has been shown that the wave function with both $d_{x^2-y^2}$ and $d_{xy}$ 
parameters could have topological order. In fact, by changing the 
boundary conditions of the BCS Hamiltonian, it should be possible to obtain 
four different projected states which in the thermodynamic limit are 
degenerate and orthogonal but, however, not connected by any local spin 
operator. In this respect, it has been argued more recently that a 
topological degeneracy may be related to the signs of the wave function and 
cannot be obtained for states satisfying the Marshall-Peierls rule~\cite{li}. 

In the spin-liquid regime, the simultaneous presence of $\Delta^{x^2-y^2}_k$ 
and $\Delta^{xy}_k$ could shift the gapless modes of the unprojected BCS 
spectrum $E_k$ from $(\pm \pi/2,\pm \pi/2)$ to incommensurate $k$-points
along the Fermi surface determined by $\epsilon_k=0$. However, we have 
demonstrated recently that a particular $\Delta^{xy}_k$ pairing, 
$\Delta^{xy}_k \propto \sin(2k_x) \sin(2k_y)$, may be imposed, in order 
to fix the nodes at the commensurate points $(\pm \pi/2,\pm \pi/2)$, 
without paying an additional energy penalty. Once $E_k$ is connected to the 
true spin excitations, a gapless spectrum is also expected. At present, 
within a pure variational technique, it is not possible to assess the 
possibility of incommensurate, gapless spin excitations being present. An 
even more challenging problem is to understand if the topological states 
could survive at all in the presence of a gapless spectrum.

\section{Other frustrated lattices}\label{sec:other}

\begin{figure}[t]
\centering
\includegraphics*[width=.45\textwidth]{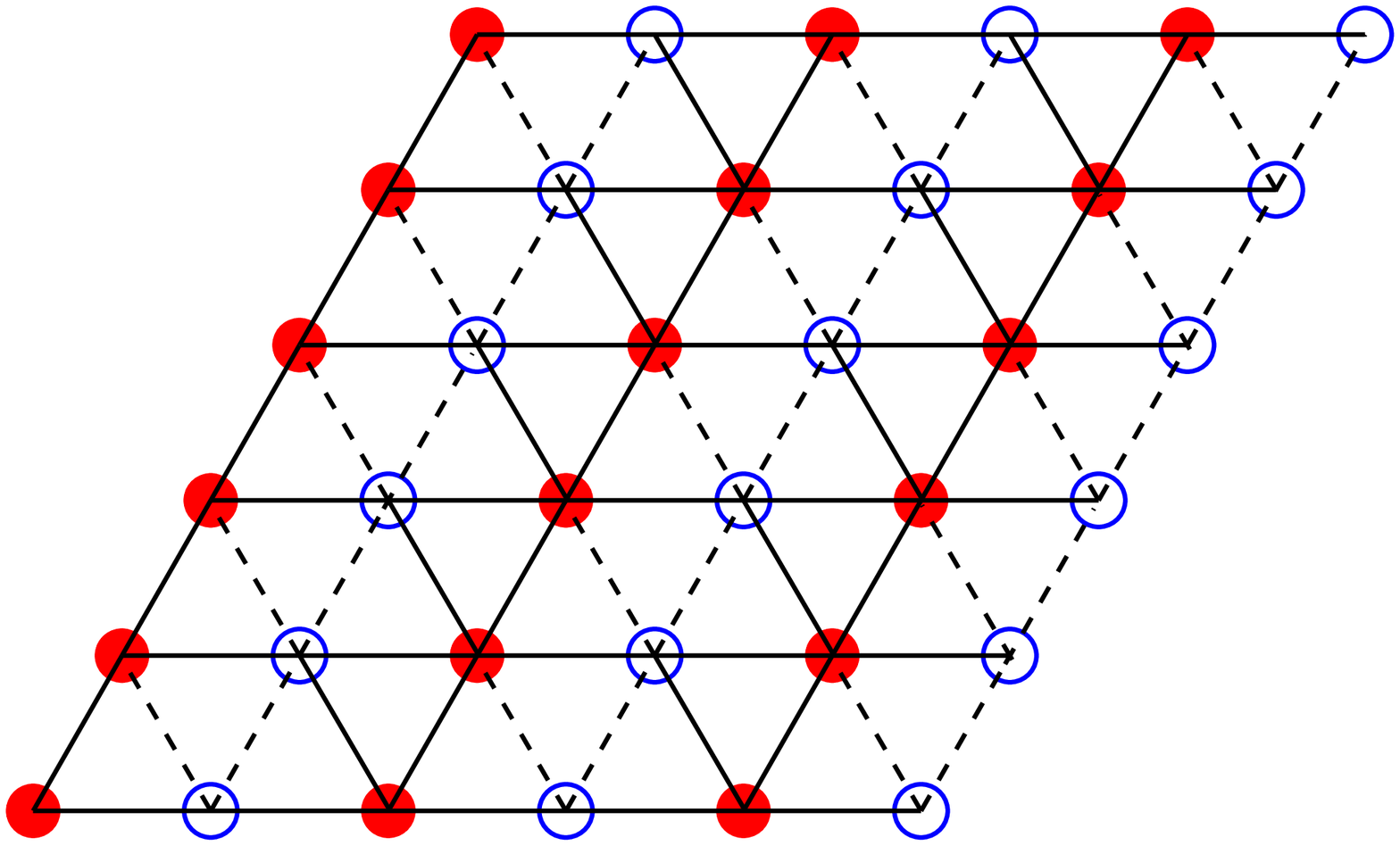}
\caption[]{Nearest-neighbor pairing function consistent with the
sign convention of the short-range RVB state in the triangular lattice: 
solid (dashed) lines represent positive (negative) values. Note that the 
unit cell contains two sites, indicated by empty and full circles.}
\label{fig:13}
\end{figure}
\begin{figure}[t]
\centering
\includegraphics*[width=.45\textwidth]{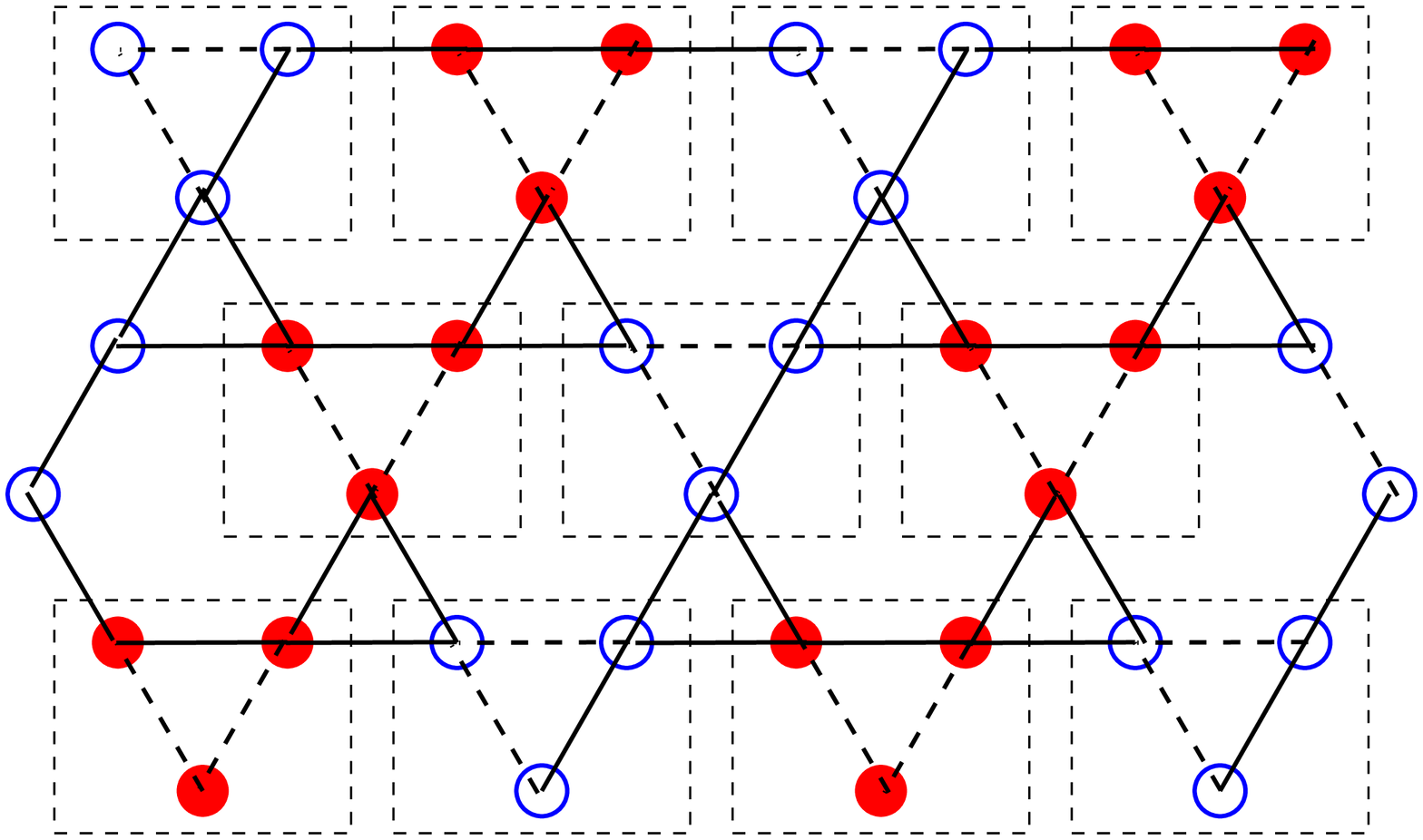}
\includegraphics*[width=.45\textwidth]{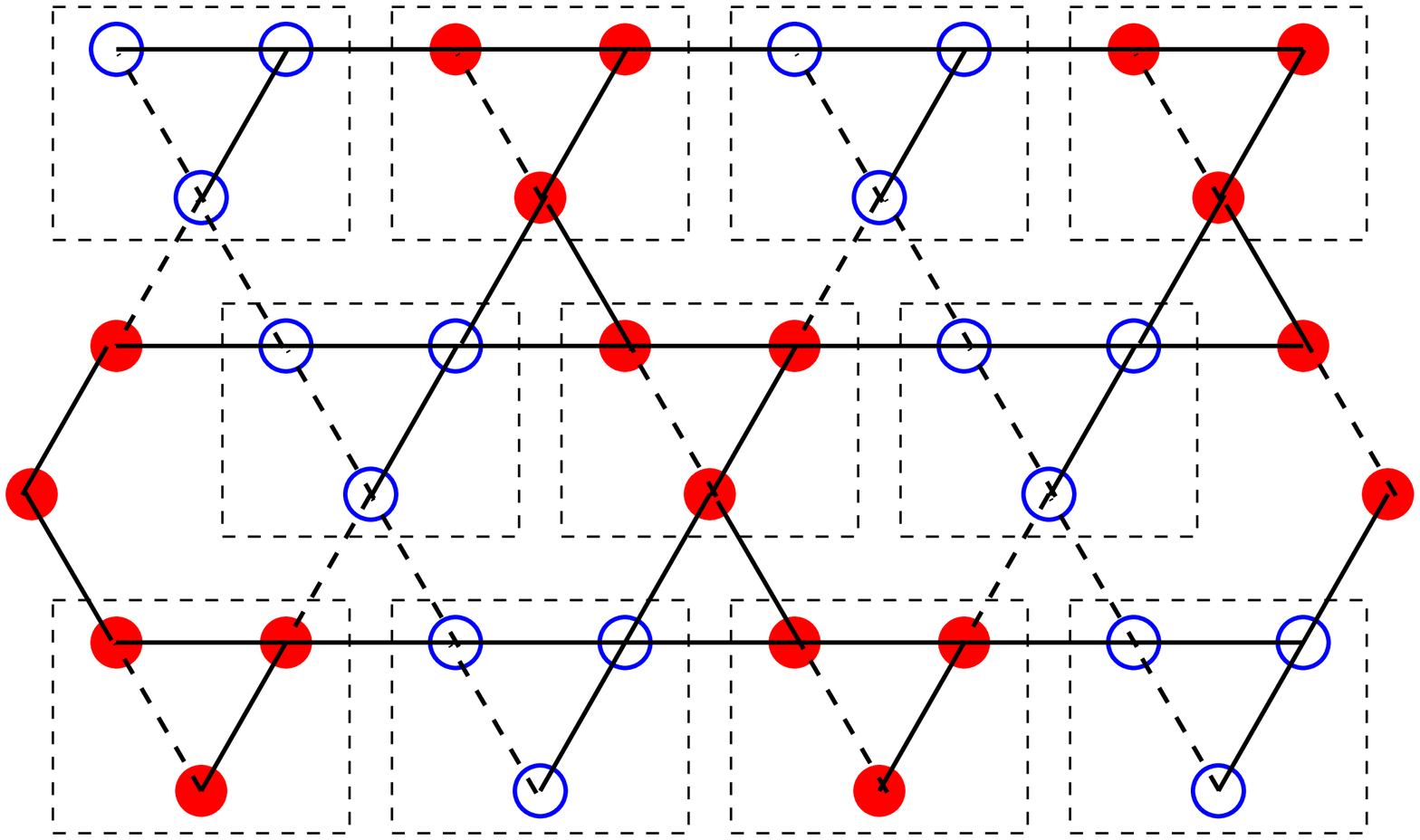}
\caption[]{Left panel: signs of the \textit{real} hopping terms 
$\chi_{R,R^\prime}$ of the U(1) Dirac spin liquid on the kagome lattice 
[Eq.~(\ref{hmfpalee})]~\cite{palee}. 
Solid (dashed) lines represent positive (negative) values. The unit cell 
contains six sites (three empty and three full circles inside the boxes). 
Right panel: nearest-neighbor pairing function consistent with the 
sign convention of the short-range RVB state in the kagome lattice: 
solid (dashed) lines represent positive (negative) values. The unit cell 
also contains six sites in this case.}
\label{fig:14-15}
\end{figure}

In this last section, we provide a brief overview of related variational 
studies performed for other lattice structures. In particular, we discuss 
in some detail the symmetries of the variational wave function on the 
anisotropic triangular lattice, considered in Ref.~\cite{yunoki}. In this 
case, one-dimensional chains with antiferromagnetic interaction $J$ are 
coupled together by a superexchange $J^\prime$, such that by varying the 
ratio $J^\prime/J$, the system interpolates between decoupled chains 
($J^\prime=0$) and the isotropic triangular lattice ($J^\prime=J$); the 
square lattice can also be described in the limit of $J=0$. The case with 
$J^\prime < J$ may be relevant for describing the low-temperature behavior 
of ${\rm Cs_2CuCl_4}$~\cite{coldea}, whereas $J^\prime \sim J$ may be 
pertinent to the insulating regime of some organic materials, such as 
$\kappa-{\rm (ET)_2Cu_2(CN)_3}$~\cite{kanoda}.

In Ref.~\cite{yunoki}, it has been shown that very accurate variational
wave functions can be constructed, providing evidence in favor of two 
different spin-liquid phases, a gapped one close to the isotropic point and
a gapless one close to the one-dimensional regime, see Fig.~\ref{fig:12}. 
We focus our attention on the isotropic point. In this case, a natural
variational \textit{ansatz} is the bosonic short-range RVB state of 
Eq.~(\ref{bosonicrvb})~\cite{sindzingre}. Exact numerical calculations for 
the $6 \times 6$ isotropic model have shown that the overlap between the 
short-range RVB wave function and the ground state is very large, 
$|\langle RVB|\Psi_0\rangle|^2=0.891$, and also that the average sign
$\langle s \rangle =\sum_x |\langle x|\Psi_0\rangle|^2 {\rm sign}  
\left\{ \langle x|\Psi_0\rangle \langle x|RVB\rangle \right\}=
0.971$~\cite{yunoki} is very close to its maximal value, $\langle s \rangle
 = 1$. We note that both the values of the overlap and of the average sign 
are much better than those obtained by a wave function that describes 
a magnetically ordered state, despite the smaller number of variational 
parameters~\cite{capriottitri}.
Although the short-range RVB state is a very good variational \textit{ansatz}, 
the bosonic representation of this state is rather difficult to handle in 
large clusters. Its systematic improvement by the inclusion of long-range
valence bonds leads to a very severe sign problem, even at the variational
level~\cite{sindzingre}. In this respect, following the rules discussed in 
section~\ref{sec:boson}, it is possible to obtain a fermionic representation 
of the short-range RVB state. The signs of the pairing function 
$f_{R,R^\prime}$ are given in Fig.~\ref{fig:13} for open boundary conditions. 
Remarkably, this particular pattern leads to a $2 \times 1$ unit cell, which 
cannot be eliminated by using local SU(2) transformations of the type 
discussed in section~\ref{sec:symm}. 
The variational RVB wave function is obtained by projecting the ground 
state of the BCS Hamiltonian, with a particular choice of the couplings: 
the only nonzero parameters are the chemical potential $\mu$ 
and the nearest-neighbor singlet gap $\Delta_{R,R^\prime}$, in the limit 
$-\mu \gg |\Delta_{R,R^\prime}|$ (so that the pairing function is proportional
to the superconducting gap). The amplitude of the gap 
$\vert \Delta_{R,R^\prime}\vert=\Delta$ is uniform, while the appropriate 
phases are shown in Fig.~\ref{fig:13}. The BCS Hamiltonian is defined on a 
$2 \times 1$ unit cell and, therefore, is not translationally invariant. 
Despite the fact that it is invariant under an elementary translation 
${\cal T}_2$ in the $\tau_2=(1/2,\sqrt{3}/2)$ direction, it is not invariant 
under an elementary translation ${\cal T}_1$ in the $\tau_1=(1,0)$ direction. 
Nevertheless, this symmetry is recovered after the projection $P_G$, 
making $|pBCS\rangle$ translationally invariant. Indeed, one can combine 
the translation operation ${\cal T}_1$ with the SU(2) gauge transformation
\begin{equation}\label{gaugetri}
c^{\dag}_{R,\sigma} \to - c^{\dag}_{R,\sigma} 
\end{equation}
for $R = m_1 \tau_1 + m_2 \tau_2 $ with $m_2$ {\em odd}. Under the composite 
application of the transformations ${\cal T}_1$ and~(\ref{gaugetri}),
the projected BCS wave function does not change. Because the gauge 
transformation acts as an identity in the physical Hilbert space with 
singly occupied sites, $|pBCS\rangle$ is translationally invariant.

\begin{table}[!pb]
\centering
\caption[]{Variational energy estimated in the thermodynamic limit
for the antiferromagnetic Heisenberg model on the isotropic triangular lattice 
($J^\prime=J$).}
\renewcommand{\arraystretch}{1.2}
\setlength\tabcolsep{5pt}
\begin{tabular}{@{}lllll@{}}
\hline\noalign{\smallskip}
wave function            & $E/J$        \\
\hline
short-range RVB          & $-0.5123(1)$ \\
RVB with $\mu=0$         & $-0.5291(1)$ \\
best RVB~\cite{yunoki}   & $-0.5357(1)$ \\
BCS+N{\'e}el~\cite{weber} & $-0.532(1)$  \\
\hline
\end{tabular}
\label{tab:2}
\end{table}

Through this more convenient representation of the short-range RVB state
by the pBCS wave function, it is possible to calculate various physical 
quantities using the standard variational Monte Carlo method. One example 
is the very accurate estimate of the variational energy per site in the 
thermodynamic limit, $E/J=-0.5123(1)$~\cite{yunoki}. Another important 
advantage of the fermionic representation is that it is easy to improve 
the variational \textit{ansatz} in a systematic way. The variational 
energy can be improved significantly by simply changing the chemical 
potential $\mu$ from a large negative value to zero, see Table~\ref{tab:2}.
We note that in this case $|pBCS\rangle$ is equivalent to a 
Gutzwiller-projected free fermion state with nearest-neighbor hoppings 
defined in a $2\times 1$ unit cell, because, through the SU(2) 
transformation of Eq~(\ref{third}), the off-diagonal pairing terms are 
transformed into kinetic terms. Further, the BCS 
Hamiltonian may be extended readily to include long-range valence bonds 
by the simple addition of nonzero $\Delta_{R,R^\prime}$ or $t_{R,R^\prime}$ 
terms. It is interesting to note that, within this approach, it is possible 
to obtain a variational energy $E/J=-0.5357(1)$ lower than that obtained by 
starting from a magnetically ordered state and considered in 
Ref.~\cite{weber}, see Table~\ref{tab:2}.

Finally, projected states have been also used to describe the ground state 
of the Heisenberg Hamiltonian on the kagome lattice~\cite{palee,palee2}. 
In this case, different possibilities for the mean-field Hamiltonian have 
been considered, with no BCS pairing but with non-trivial fluxes through 
the triangles and the hexagons of which the kagome structure is composed. 
In particular, the best variational state in this class can be found by 
taking
\begin{equation}\label{hmfpalee}
{\cal H}_{MF} = - \sum_{\langle R,R^\prime \rangle,\sigma} 
\chi_{R,R^\prime} c^\dag_{R,\sigma} c_{R^\prime,\sigma} + H.c.,
\end{equation}
with all the hoppings $\chi_{R,R^\prime}$ having the same magnitude and 
producing a zero flux through the triangles and $\pi$ flux through the 
hexagons. One may fix a particular gauge in which all $\chi_{R,R^\prime}$ 
are real, see Fig.~\ref{fig:14-15}. In this gauge, the mean-field spectrum has 
Dirac nodes at $k=(0,\pm \pi/\sqrt{3})$, and the variational state describes a 
U(1) Dirac spin liquid. Remarkably, this state should be stable against 
dimerization (i.e., it has a lower energy than simple valence-bond
solids), in contrast to mean-field results~\cite{hastings2}. Another 
competing mean-field state~\cite{hastings2}, which is obtained by 
giving the fermions chiral masses and is characterized by a broken 
time-reversal symmetry (with $\theta$ flux through triangles and 
$\pi-\theta$ flux through hexagons), is also found to have a higher 
energy than the pure spin-liquid state. In this context, it would be 
valuable to compare the wave function proposed in Ref.~\cite{palee} 
with the systematic improvement of the short-range RVB state which 
has a simple fermionic representation (see Fig.~\ref{fig:14-15}).

\section{Conclusions}\label{sec:conc}

In summary, we have shown that projected wave functions containing both
electronic pairing and magnetism provide an extremely powerful tool
to study highly frustrated magnetic materials. In particular, these pBCS
states may describe all known phases in one-dimensional systems, giving
very accurate descriptions when compared to state-of-the-art DMRG calculations.
Most importantly, variational wave functions may be easily generalized to
treat higher dimensional systems: here we have presented in detail the
case of the two-dimensional $J_1{-}J_2$ model, as well as some examples
of other frustrated lattices which have been considered in the recent past.

The great advantage of this variational approach in comparison with other
methods, such as DMRG, is that it can offer a transparent description of
the ground-state wave function. Furthermore, the possibility of giving a
physical interpretation of the \textit{unprojected} BCS spectrum $E_k$,
which is expected to be directly related to the true spin excitations,
is very appealing. We demonstrated that this correspondence works very
well in one dimension, both for gapless and for dimerized phases. In two
dimensions, the situation is more complicated and we close by expressing
the hope that future investigations may shed further light one the
fascinating world of the low-energy properties of disordered magnetic
systems.

\section*{Acknowledgments}

We have had the privilege of discussing with many people over the lifetime 
of this project, and would like to express our particular thanks to 
P. Carretta, D. Ivanov, P.A. Lee, C. Lhuillier, F. Mila, G. Misguich, 
D. Poilblanc, A.W. Sandvik, and X.-G. Wen. We also thank M. Mambrini and 
K.S.D. Beach for providing us with the energies of the bosonic RVB wave 
function in Table~\ref{tab:1}, A.W. Sandvik for the bosonic data shown in 
Fig.~\ref{fig:5-6}, and S.R. White for the DMRG data shown in 
Figs.~\ref{fig:1-2} and ~\ref{fig:3-4}. We acknowledge partial support 
from CNR-INFM.

%
%

%

\begin{thebibliography}{99.}

\bibitem{bcs} J.R. Schrieffer, ``Theory of Superconductivity'', Addison 
   Wesley (1964).
\bibitem{laughlin} R.B. Laughlin, Phys. Rev. Lett. \textbf{50}, 1395 (1983).
\bibitem{fazekas} P.W. Anderson, 
   Mater. Res. Bull \textbf{8}, 153 (1973); 
   P. Fazekas and P.W. Anderson, 
   Philos. Mag. \textbf{30}, 423 (1974).
\bibitem{misguich} For a review see 
   G. Misguich and C. Lhuillier,
   in ``Frustrated Spin Models'', Ed. H. T. Diep, World Scientific, New Jersey
   (2004); see also G. Misguich in this volume.
\bibitem{castilla} G. Castilla, S. Chakravarty, and V.J. Emery,
   Phys. Rev. Lett. \textbf{75}, 1823 (1995).
\bibitem{carretta} R. Melzi, P. Carretta, A. Lascialfari, M. Mambrini,
   M. Troyer, P. Millet, and F. Mila, 
   Phys. Rev. Lett. \textbf{85}, 1318 (2000).
\bibitem{carretta2} P. Carretta, N. Papinutto, C. B. Azzoni, M. C. Mozzati, 
   E. Pavarini, S. Gonthier, and P. Millet, 
   Phys. Rev. B \textbf{66}, 094420 (2002).
\bibitem{white} S.R. White and I. Affleck,
   Phys. Rev. B \textbf{54}, 9862 (1996).
\bibitem{eggert} S. Eggert, 
   Phys. Rev. B \textbf{54}, 9612 (1996).
\bibitem{majumdar1} C.K. Majumdar and D.K. Ghosh, 
   J. Math. Phys. \textbf{10}, 1388, (1969). 
\bibitem{majumdar2} C.K. Majumdar and D.K. Ghosh, 
   J. Math. Phys. \textbf{10}, 1399 (1969). 
\bibitem{doucot} P. Chandra and B. Doucot, 
   Phys. Rev. B \textbf{38}, 9335 (1988).
\bibitem{dagotto} E. Dagotto and A. Moreo,
   Phys. Rev. Lett. \textbf{63}, 2148 (1989).
\bibitem{singh} R.R.P. Singh and R. Narayanan,
   Phys. Rev. Lett. \textbf{65}, 1072 (1990).
\bibitem{schulz} J. Schulz, T.A. Ziman, and D. Poilblanc,
   J. Phys. I \textbf{6}, 675 (1996).
\bibitem{gelfand} M.P. Gelfand, R.R.P. Singh, and D.A. Huse,
   Phys. Rev. B \textbf{40}, 10801 (1989).
\bibitem{singh2} R.R.P. Singh, Z. Weihong, C.J. Hamer, and J. Oitmaa,
   Phys. Rev. B \textbf{60}, 7278 (1999).
\bibitem{kotov} V.N. Kotov, J. Oitmaa, O.P. Sushkov, and Z. Weihong, 
   Phil. Mag. B \textbf{80}, 1483 (2000).
\bibitem{sushkov} O.P. Sushkov, J. Oitmaa, and Z. Weihong, 
   Phys. Rev. B \textbf{63}, 104420 (2001).
\bibitem{read} N. Read and S. Sachdev, 
   Phys. Rev. Lett. \textbf{62}, 1694 (1989).
\bibitem{mambrini} M. Mambrini, A. Lauchli, D. Poilblanc, and F. Mila,
   Phys. Rev. B \textbf{74}, 144422 (2006).
\bibitem{lieb} E. Lieb and D. Mattis, 
   J. Math. Phys. \textbf{3}, 749 (1962).
\bibitem{marshall} W. Marshall, 
   Proc. R. Soc. London Ser. \textbf{A 232}, 48 (1955).
\bibitem{richter} J. Richter, N.B. Ivanov, and K. Retzlaff, 
   Europhys. Lett. \textbf{25}, 545 (1994).
\bibitem{ceperley} D.F.B. ten Haaf, H.J.M. van Bemmel, J.M.J. van Leeuwen, 
   W. van Saarloos, and D.M. Ceperley, 
   Phys. Rev. B \textbf{51}, 13039 (1995).
\bibitem{anderson2} P.W. Anderson, 
   Science \textbf{235}, 1196 (1987).
\bibitem{liang} S. Liang, B. Doucot, and P.W. Anderson, 
   Phys. Rev. Lett. \textbf{61}, 365 (1988).
\bibitem{sindzingre} P. Sindzingre, P. Lecheminant, and C. Lhuillier,
   Phys. Rev. B \textbf{50}, 3108 (1994).
\bibitem{becca} F. Becca, L. Capriotti, A. Parola, and S. Sorella,
   Phys. Rev. B \textbf{76}, 060401 (2007).
\bibitem{affleck} I. Affleck, Z. Zou, T. Hsu, and P.W. Anderson, 
   Phys. Rev. B \textbf{38}, 745 (1988).
\bibitem{rice} F.-C. Zhang, C. Gros, T.M. Rice, and H. Shiba,
   Supercond. Sci. Technol. \textbf{36}, 1 (1988).
\bibitem{wen} X.-G. Wen, 
   Phys. Rev. B \textbf{65}, 165113 (2002).
\bibitem{haldane} F.D.M. Haldane, 
   Phys. Rev. Lett. \textbf{60}, 635 (1988).
\bibitem{lou} J. Lou and A.W. Sandvik,
   Phys. Rev. B \textbf{76}, 104432 (2007).
\bibitem{sandvik} K.S.D. Beach and A.W. Sandvik,
   Nucl. Phys. B \textbf{750}, 142 (2006).
\bibitem{sandvik2} A.W. Sandvik and K.S.D. Beach,
   arXiv:0704.1469.
\bibitem{beach} K.S.D. Beach,
   arXiv:0709.3297.
\bibitem{sutherland} B. Sutherland,
   Phys. Rev. B \textbf{37}, 3786 (1988).
\bibitem{chakraborty} N. Read and B. Chakraborty,
   Phys. Rev. B \textbf{40}, 7133 (1989).
\bibitem{kasteleyn}  P.W. Kasteleyn, 
   J. Math. Phys. \textbf{4}, 287 (1963).
\bibitem{gros} C. Gros,
   Phys. Rev. B \textbf{42}, 6835 (1990).
\bibitem{manousakis} E. Manousakis, 
   Rev. Mod. Phys. \textbf{63}, 1 (1991).
\bibitem{franjio} F. Franjic and S. Sorella, 
   Prog. Theor. Phys. \textbf{97}, 399 (1997).
\bibitem{bouchaud} J.P. Bouchaud, A. Georges, and C. Lhuillier, 
   J. Phys. (Paris) \textbf{49}, 553 (1988).
\bibitem{sorella} S. Sorella, 
   Phys. Rev. B \textbf{71}, 241103 (2005).
\bibitem{gros2} C. Gros,
   Ann. of Phys. \textbf{189}, 53 (1989).
\bibitem{affleck2} I. Affleck, D. Gepner, H.J. Schulz, and T. Ziman,
   J. Phys. A \textbf{22}, 511 (1989).
\bibitem{sorella2} S. Sorella, L. Capriotti, F. Becca, and A. Parola,
   Phys. Rev. Lett. \textbf{91}, 257005 (2003).
\bibitem{parola} A. Parola, S. Sorella, F. Becca, and L. Capriotti,
   condmat/0502170.
\bibitem{reger} J.D. Reger and A.P. Young,
   Phys. Rev. B \textbf{37}, 5978 (1988).
\bibitem{sandvik3} A.W. Sandvik,
   Phys. Rev. B \textbf{56}, 11678 (1997).
\bibitem{calandra} M. Calandra Buonaura and S. Sorella,
   Phys. Rev. B \textbf{57}, 11446 (1998).
\bibitem{sandvik4} A.W. Sandvik, private communication.
\bibitem{capriotti} L. Capriotti, F. Becca, A. Parola, and S. Sorella,
   Phys. Rev. Lett. \textbf{87}, 097201 (2001).
\bibitem{mila} F. Mila, D. Poilblanc, and C. Bruder,
   Phys. Rev. B \textbf{43}, 7891 (1991).
\bibitem{darradi} R. Darradi, O. Derzhko, R. Zinke, J. Schulenburg, 
   S.E. Krueger, and J. Richter,
   arXiv:0806.3825.
\bibitem{mambrini2} M. Mambrini, private communication.
\bibitem{beach2} K.S.D. Beach, private communication.
\bibitem{capriotti2} L. Capriotti, unpublished.
\bibitem{hastings} M.B. Hastings,
   Phys. Rev. B \textbf{69}, 104431 (2004).
\bibitem{lsm} E.H. Lieb, T.D. Schultz, and D.C. Mattis, 
   Ann. Phys. (N.Y.) \textbf{16}, 407 (1961).
\bibitem{ivanov} D.A. Ivanov and T. Senthil,
   Phys. Rev. B \textbf{66}, 115111 (2002).
\bibitem{li} T. Li and H.-Y. Yang,
   Phys. Rev. B \textbf{75}, 172502 (2007).
\bibitem{yunoki} S. Yunoki and S. Sorella, 
   Phys. Rev. B \textbf{74}, 014408 (2006).
\bibitem{coldea} R. Coldea, D.A. Tennant, A.M. Tsvelik, and Z. Tylczynski,
   Phys. Rev. Lett. \textbf{86}, 1335 (2001).
\bibitem{kanoda} Y. Shimizu, K. Miyagawa, K. Kanoda, M. Maesato, and G. Saito,
   Phys. Rev. Lett. \textbf{91}, 107001 (2003).
\bibitem{capriottitri} L. Capriotti, A.E. Trumper, and S. Sorella,
   Phys. Rev. Lett. \textbf{82}, 3899 (1999).
\bibitem{weber} C. Weber, A. Laeuchli, F. Mila, and T. Giamarchi, 
   Phys. Rev. B \textbf{73}, 014519 (2006).
\bibitem{palee} Y. Ran, M. Hermele, P.A. Lee, and X.-G. Wen,
   Phys. Rev. Lett. \textbf{98}, 117205 (2007).
\bibitem{palee2} M. Hermele, Y. Ran, P.A. Lee, and X.-G. Wen,
   Phys. Rev. B \textbf{77}, 224413 (2008).
\bibitem{hastings2} M.B. Hastings,
   Phys. Rev. B \textbf{63}, 014413 (2000).

\end{thebibliography}
%

\printindex
\end{document}